\documentclass[11pt,a4paper]{article}
\pdfoutput=1
\usepackage{jheppub}

\usepackage{epsfig}
\usepackage{graphicx}
\usepackage{subfig}
\usepackage{amssymb}
\usepackage{amsmath}
\usepackage{booktabs,multirow,tabularx}
\usepackage{lmodern}
\usepackage{bigdelim}
\usepackage{slashed}
\usepackage{caption}
\usepackage{float}
\usepackage{placeins}
\usepackage{rotating}
\usepackage{lscape}
\usepackage{listings}
\usepackage{comment}
\usepackage[Q=yes,pverb-linebreak=no]{examplep}

\usepackage{morefloats}
\usepackage{pgffor}
\usepackage{array}
\usepackage{lineno}

\DeclareGraphicsExtensions{.eps}
\graphicspath{{./}}

\newcommand{\LHCbPlotDir}{./Figures/LHCb}
\newcommand{\ATLASPlotDir}{./Figures/ATLAS}
\newcommand{\CMSPlotDir}{./Figures/CMS}

\def\beq{\begin{equation}}
\def\beqn{\begin{eqnarray}}
\def\eeq{\end{equation}}
\def\eeqn{\end{eqnarray}}
\def\beal{\begin{align}}
\def\endal{\end{align}}

\newcommand{\ie}{{\it i.e.}}
\newcommand{\eg}{{\it e.g.}}

\newcommand{\ben}{\begin{enumerate}}
\newcommand{\een}{\end{enumerate}}
\newcommand{\bit}{\begin{itemize}}
\newcommand{\eit}{\end{itemize}}
\newcommand{\bc}{\begin{center}}
\newcommand{\ec}{\end{center}}
\newcommand{\bq}{\begin{equation}}
\newcommand{\eq}{\end{equation}}
\newcommand{\bqa}{\begin{eqnarray}}
\newcommand{\eqa}{\end{eqnarray}}

\def\MG5aMC{{\sc \small MadGraph5\_aMC@NLO}}

\title{Prompt ${J/\psi}$-pair production at the LHC: impact of loop-induced contributions and of the colour-octet mechanism}

\author[a]{Jean-Philippe Lansberg,}

\author[b]{Hua-Sheng Shao,}

\author[a,c]{Nodoka Yamanaka,}

\author[d,e]{Yu-Jie Zhang}

\affiliation[a]{IPNO, CNRS-IN2P3, Univ. Paris-Sud, Universit\'e Paris-Saclay, 
91406 Orsay Cedex, France}
\affiliation[b]{Laboratoire de Physique Th\'eorique et Hautes Energies (LPTHE), UMR 7589, Sorbonne Universit\'e et CNRS, 4 place Jussieu, 75252 Paris Cedex 05, France}
\affiliation[c]{Yukawa Institute for Theoretical Physics, Kyoto University, Kitashirakawa-Oiwake, Kyoto 606-8502, Japan}
\affiliation[d]{School of Physics and Nuclear Energy Engineering, Beihang University, Beijing 100083, China}
\affiliation[e]{Center for High Energy Physics, Peking University, Beijing 100871, China}

\emailAdd{Jean-Philippe.Lansberg@in2p3.fr}
\emailAdd{huasheng.shao@lpthe.jussieu.fr}
\emailAdd{yamanaka@ipno.in2p3.fr}
\emailAdd{nophy0@gmail.com}

\abstract{Prompt double-$J/\psi$ production at high-energy hadron colliders can be considered as a golden channel to probe double parton scatterings (DPS) --in particular to study gluon-gluon correlations inside the proton-- and, at the same time, to measure the distribution of linearly-polarised gluons inside the proton. Such studies however require a good control of both single and DPS in the respective regions where they are carried out. In this context, we have critically examined two mechanisms of single parton scatterings (SPS) that may be kinematically enhanced
where DPS are thought to be dominant, 
even though they are either at higher orders in the strong-coupling or velocity expansion.
First, we have considered 
 a gauge-invariant and infrared-safe subset of the loop-induced contribution via Colour-Singlet (CS) transitions. We have found it to become the leading CS SPS contributions at large rapidity separation, yet too small to account for the data without 
invoking the presence of DPS yields. Second, we have surveyed the possible Colour-Octet (CO) contributions using both old and up-to-date non-perturbative long distance matrix elements (LDMEs). We have found that the pure CO yields crucially depend on the LDMEs. Among all the LDMEs we used, only two result into a visible modification of the NRQCD (CS+CO) yield, but only in two kinematical distributions measured by ATLAS, those of the rapidity separation and of the pair invariant mass. These modifications however do not impact the control region used for their DPS study.
}


\date{\today}

\begin{document}

\maketitle

%

 
 
 


\section{Introduction}

The role of multiple parton interactions in proton-proton collisions is believed to become increasingly important when one explores the energy frontier in particle physics. As such, the relevance in LHC observables  of two simultaneous hard scatterings, usually referred to as Double Parton Scatterings (DPS), has attracted much attention in the last decade with significant theory advances related to perturbative QCD~\cite{Blok:2011bu,Diehl:2011yj,Diehl:2011tt,Gaunt:2011xd,Manohar:2012pe,Gaunt:2012dd,Blok:2013bpa,Diehl:2014vaa,Diehl:2015bca,Rinaldi:2016jvu,Buffing:2017mqm,Diehl:2017kgu,Vladimirov:2017ksc,Diehl:2018wfy,Gaunt:2018eix}. Since DPS are higher-twist effects in total cross sections compared to the conventional single parton scatterings (SPS), quantitative studies of DPS remain challenging though not impossible both on the theoretical and experimental sides. These are particularly interesting since they provide us with means to study parton correlations inside the proton (see \eg\ \cite{Rinaldi:2016jvu,Rinaldi:2018bsf,Rinaldi:2018slz}).


Among the possible hard probes of DPS at high-energy hadron colliders, the associated production of quarkonia (see~\cite{Lansberg:2019adr} for an exhaustive review) provides unique opportunities to measure DPS in gluon-induced reactions thus to study gluon-gluon correlations in the proton. Numerous measurements of quarkonium associated processes have been performed at the Tevatron and the LHC. They can mainly be categorised as di-quarkonium production ($J/\psi+J/\psi$~\cite{Aaij:2011yc,Abazov:2014qba,Khachatryan:2014iia,Aaboud:2016fzt,Aaij:2016bqq}, $J/\psi+\Upsilon$~\cite{Abazov:2015fbl}, $\Upsilon+\Upsilon$~\cite{Khachatryan:2016ydm}), associated production with a vector boson ($J/\psi+W^{\pm}$~\cite{Aad:2014rua}, $J/\psi+Z$~\cite{Aad:2014kba}) or with another heavy quark  ($J/\psi+$open charm~\cite{Aaij:2012dz}, $\Upsilon+$open charm~\cite{Aaij:2015wpa}). All these measurements cover different kinematical regions with different momentum transfers in the hard scattering. Their theoretical analysis is highly non-trivial, which has triggered many theoretical studies in the recent years~\cite{Li:2009ug,Qiao:2009kg,Ko:2010xy,Kom:2011bd,Berezhnoy:2011xy,Lansberg:2013qka,Li:2013csa,Lansberg:2014swa,Sun:2014gca,Lansberg:2015lva,He:2015qya,Baranov:2015cle,Shao:2016wor,Lansberg:2016rcx,Lansberg:2016muq,Likhoded:2016zmk,Borschensky:2016nkv,Lansberg:2017chq,Lansberg:2017dzg,Cisek:2017ikn,Gridin:2019nhc}.  Very recently, the first calculation of triple-$J/\psi$ production showed that it can help us probe both DPS and triple parton scatterings (TPS)~\cite{Shao:2019qob}.

In this context, we focus in this paper on the di-$J/\psi$ case with the aim to improve the existing perturbative QCD calculations for the SPS. To do so, we consider higher-order corrections in both the strong coupling constant, $\alpha_S$, and the heavy-quark velocity, $v$. First, we study the impact of a gauge-invariant and infrared-safe subset of loop-induced (LI) contributions. Our analysis follows the lines of a similar study for $J/\psi+\Upsilon$ production~\cite{Shao:2016wor}. Such contributions appear  at next-to-next-to-leading order (NNLO) in $\alpha_S$ but could be enhanced at large rapidity differences and high invariant masses of the $J/\psi$ pair because of the presence of topologies with double $t$-channel gluon exchanges  between both charm-anticharm quark lines. Second, we perform a comprehensive survey of the impact of the colour-octet (CO) contributions in three kinematical domains covered by the existing LHC measurements~\cite{Khachatryan:2014iia,Aaboud:2016fzt,Aaij:2016bqq} considering the various existing fits of the non-perturbative CO long-distance matrix elements (LDMEs).

In order to disentangle DPS from SPS in observables where two particles are observed, one usually relies on the analysis of specific kinematical dependences which are believed to be drastically different in both samples. Common choices of variables are the azimuthal and the rapidity separations between both observed particles, $\Delta \phi$ and $\Delta y$. 
The DPS contribution, coming from two {\it a priori} independent parton scatterings, is expected to be flatter than the SPS one in both distributions.

For double-$J/\psi$ studies, the analysis of the $\Delta y(J/\psi,J/\psi)$ distributions should be preferred compared to that of $\Delta \phi(J/\psi,J/\psi)$ since the $\Delta y(J/\psi,J/\psi)$ distribution of the SPS yield is much less affected by possible non-perturbative intrinsic $k_T$ of the colliding gluons~\cite{Kom:2011bd} than the $\Delta \phi(J/\psi,J/\psi)$ one, which can become as flat as the DPS ones in some cases. In general, one expects the DPS fraction to be the largest at large $|\Delta y(J/\psi,J/\psi)|$. A precise determination of the DPS yield therefore requires a good knowledge of the SPS in this region. Both the LI and CO topologies with $t$-channel-gluon exchanges could result into a flat $\frac{d\sigma}{d\Delta y}$ like in the $J/\psi+\Upsilon$ case~\cite{Shao:2016wor}. 

Assuming $\alpha_S\sim v^2$, the colour-singlet (CS) LI contribution should be of the same magnitude as the leading order (LO) CO contribution (yet both smaller that the bulk of the CS yield in the absence of the possible kinematical enhancement which we are after here). According to the NRQCD velocity scaling rules~\cite{Bodwin:1994jh}, the former one is indeed $\mathcal{O}(\alpha_S^6v^3)$ while the latter one is $\mathcal{O}(\alpha_S^4v^7)$. This justifies why we consider both of them in this study.

This article is organised as follows. In section \ref{sec:lhcmeasure}, we first quickly review the existing LHC measurements used in our comparisons~\footnote{We do not consider the D0 measurement~\cite{Abazov:2014qba} at the Tevatron since no corrected distribution was released which could be used for a direct data-theory comparison.}. Then, we discuss our theory framework in section~\ref{sec:theory}. Section \ref{sec:lipart} gathers our discussion of the impact of the inclusion LI CS contribution and section \ref{sec:copart} comprises a comprehensive analysis of complete LO CO contribution. The appendix \ref{app:moreplots} collects additional plots relevant for further theory-data comparisons.

\section{LHC measurements and kinematical variables\label{sec:lhcmeasure}}

\subsection{kinematical variables\label{app:variables}}

We start by introducing the kinematical variables relevant for di-quarkonium production. On the experimental side, the second LHCb analysis~\cite{Aaij:2016bqq} bears on the largest set of the kinematical variables whose distribution is used for comparisons between the experimental measurements and the theoretical calculations. Since some of these variables may not be very common, we summarise the description of their names or labels in Table~\ref{tab:variables}. In particular, the transverse momentum asymmetry is defined as
\begin{eqnarray}
A_{T}(J/\psi,J/\psi)\equiv \left|\frac{P_T(J/\psi_1)-P_T(J/\psi_2)}{P_T(J/\psi_1)+P_T(J/\psi_2)}\right|,
\end{eqnarray}
where $J/\psi_1$ and $J/\psi_2$ are respectively denoted as the first and second hardest $J/\psi$ with ordered in the transverse momentum. 

\begin{table}[ht]
\begin{center}
\begin{tabular}{c||c}
\hline\hline
$P_T(J/\psi+J/\psi)$ & $y(J/\psi+J/\psi)$\\\hline
Transverse momentum of the pair & Rapidity of the pair \\
\hline\hline
$P_T(J/\psi)$ & $y(J/\psi)$ \\\hline
Transverse momentum of a randomly chosen $J/\psi$  & Rapidity of a randomly chosen $J/\psi$\\\hline\hline
$\Delta \phi(J/\psi,J/\psi)$ & $\Delta y(J/\psi,J/\psi)$\\\hline
Azimuthal angle difference in the transverse plane & Rapidity separation \\\hline\hline
$M(J/\psi+J/\psi)$ & $A_{T}(J/\psi,J/\psi)$ \\\hline
Invariant mass of the pair & Transverse momentum asymmetry \\\hline
\hline
\end{tabular}
\caption{Summary of the kinematical variables.}\label{tab:variables}
\end{center}
\end{table}

\subsection{Available data sets}

Four LHC studies of double prompt $J/\psi$ production have so far been performed~\cite{Aaij:2011yc,Khachatryan:2014iia,Aaboud:2016fzt,Aaij:2016bqq}. LHCb performed two measurements in the same kinematical region, one at $\sqrt{s}=7$ TeV and another at $\sqrt{s}=13$ TeV; we will focus on the latter which is more precise~\cite{Aaij:2016bqq}. The various kinematical cuts used in the ATLAS, CMS and LHCb analyses are summarised in Table.~\ref{tab:cuts} along with the corresponding centre-of-mass energy $\sqrt{s}$ . It is useful to note that due to the different trigger and acceptance constraints on the ATLAS, CMS and LHCb data taking, the 3 samples cover complementary domains in $P_T$ and $y$. In particular, ATLAS~\cite{Aaboud:2016fzt} imposes the largest $P_T(J/\psi)$ cut (as large as 8.5 GeV), while LHCb~\cite{Aaij:2016bqq} does not impose any lower $P_T$ cut on the observed $J/\psi$. As such, LHCb events are mostly located at low $P_T(J/\psi)$. CMS~\cite{Khachatryan:2014iia} imposes varying cuts from $P_T(J/\psi)>4.5$~GeV to $P_T(J/\psi)>6.5$~GeV depending on the rapidity. Moreover, LHCb can only detect forward particles whereas ATLAS/CMS have a generally larger rapidity coverage but in the central-rapidity region. In section \ref{sec:copart}, we will discuss how these kinematical coverages can be relevant to determine the proper CO LDMEs.

\begin{table}[ht]
\begin{center}\small	
\begin{tabular}{c|c|c}
Experiment & $\sqrt{s}$ [TeV] & Kinematical cuts\\\hline\hline
\multirow{3}{*}{CMS~\cite{Khachatryan:2014iia}} & \multirow{3}{*}{$7$} & $P_T(J/\psi)>$6.5 GeV when $|y(J/\psi)|<1.2$;\\
& & $P_T(J/\psi)>6.5-\frac{200}{23}(|y(J/\psi)|-1.2)$ GeV when $1.2<|y(J/\psi)|<1.43$;\\
& & $P_T(J/\psi)>4.5$ GeV when $1.43<|y(J/\psi)|<2.2$ \\\hline
\multirow{3}{*}{ATLAS~\cite{Aaboud:2016fzt}} & \multirow{3}{*}{$8$} & $P_T(\mu)>$2.5 GeV, $|\eta(\mu)|<2.3$;\\
& & One $J/\psi$ has two muons with $P_T(\mu)>$ 4 GeV; \\
& & $P_T(J/\psi)>$ 8.5 GeV, $|y(J/\psi)|<2.1$\\\hline
LHCb~\cite{Aaij:2016bqq} & $13$ & $P_T(J/\psi)<14$ GeV, $2.0<y(J/\psi)<4.5$ \\
\hline
\end{tabular}
\caption{Summary of kinematical cuts of the double-$J/\psi$ measurements by the LHC experiments which we will consider here.}\label{tab:cuts}
\end{center}
\end{table}

\section{Theory framework}
\label{sec:theory}
In this section, we briefly address some specificities of our theoretical computations, which however remain very standard.

\subsection{Intrinsic initial-$k_T$ smearing\label{sec:ktsmear}}

An important effect for an accurate description of double-$J/\psi$ hadroproduction is known to be the smearing of the kinematics  arising from the intrinsic $k_T$ of the gluons~\cite{Sridhar:1998rt}. It is in principle a non-perturbative effect which cannot properly be accounted for by the collinear factorisation. In fact, double-$J/\psi$ production can provide new insights in the transverse dynamics of the gluons as it was shown~\cite{Lansberg:2017dzg} using the transverse-momentum dependent (TMD) factorisation. Clearly, a collinear computation is not meant to encapsulate such effects. As a makeshift, we simply rely on an empirical procedure to deal with them which we believe to be sufficient for our phenomenological purpose. In particular, the whole $k_T$ smearing is assumed to be factorised out by
\begin{eqnarray}
\frac{d\sigma}{d\Phi_{\langle k_T \rangle}}=\int_{0}^{+\infty}{dk_T^2 \frac{\pi}{8\langle k_T\rangle^2}e^{-\frac{\pi}{8}\frac{k_T^2}{\langle k_T\rangle^2}}\frac{d\sigma}{d\Phi}},
\end{eqnarray} 
where the phase-space mapping $\Phi\rightarrow \Phi_{\langle k_T \rangle}$ is determined by boosting the whole event according to the generated transverse-momentum imbalance $|\overrightarrow{k_T}|=k_T$ with a uniform distribution of the azimuthal angle in the transverse plane. Other forms are of course possible. 
In the present study, we assume $\langle k_T \rangle$ to be the same for all three experimental coverages and fix its value to be $3.0$ GeV. The distributions with other $\langle k_T \rangle$ values are also  not shown but can easily be obtained with the help of {\sc\small HELAC-Onia}~\cite{Shao:2012iz,Shao:2015vga}. In fact, the NLO$^\star$ distributions with $\langle k_T \rangle=0.5$ GeV and $2.0$ GeV can be found in a theory-data comparison made by LHCb~\cite{Aaij:2016bqq}. The $k_T$-smearing effect is only visible for the $P_T(J/\psi+J/\psi)$, $\Delta \phi(J/\psi,J/\psi)$ and $A_T(J/\psi,J/\psi)$ distributions.

\subsection{Parameters entering our calculations\label{sec:setup}}

We now quickly describe our set-up for the present calculations before discussing the numerical results. We have fixed the charm quark mass to be 1.5 GeV and only the light $u$, $d$, $s$ (anti)quarks and the gluons are allowed in the initial states. In order to be compatible with our previous NLO$^\star$ calculations, we have used the NLO parton-distribution functions (PDFs) CTEQ6M~\cite{Pumplin:2002vw} for the calculations in the ATLAS and CMS acceptances and NNPDF3.0~\cite{Ball:2014uwa} for those in the LHCb acceptance. We have explicitly checked that the PDF dependence is less than $20\%$ and is thus a minor source of uncertainty compared to the (dominant) scale uncertainty which we discuss below. The missing higher-order terms in $\alpha_S$ are  estimated in the usual way by independently varying the factorisation and renormalisation scales  as $(\mu_F,\mu_R)=(\zeta_1\mu_0,\zeta_2\mu_0)$, with $\zeta_{1,2}=\frac{1}{2},1,2$, where the central scale $\mu_0$ is chosen to be $\mu_0=\sqrt{\left(P_T(J/\psi)\right)^2+ (4 m_c)^2}$, like in Refs.~\cite{Lansberg:2013qka,Lansberg:2014swa}. The CS LDME is estimated via
$\langle\mathcal{O}^{H_{Q\bar{Q}}}(^3S_1^{[1]})\rangle=2N_c\frac{3}{4\pi}\left|R^{H_{Q\bar{Q}}}(0)\right|^2$, where the wave function at the origin $R^{H_{Q\bar{Q}}}(0)$ can be determined by solving the Schr{\"o}dinger equation with a given QCD potential. We will use the numerical values $\left|R^{J/\psi}(0)\right|^2=0.8$ GeV$^3$ and $\left|R^{\psi(2S)}(0)\right|^2=0.5$ GeV$^3$ derived in Ref.~\cite{Eichten:1995ch} using the QCD-motivated Buchm{\"u}ller-and-Tye potential~\cite{Buchmuller:1980su}. For the CS SPS yield, the feed-down contribution from the $\psi(2S)$ decays is as large as the direct double $J/\psi$ production. In practice, we take it to be equal to 2.  It is thus mandatory to take it into account.

\section{Colour-singlet contributions: partial loop-induced corrections\label{sec:lipart}}

In principle, considering the square of a one-loop amplitude by itself should give divergent results from both the infrared and ultraviolet regions. Such one-loop amplitudes squared are part of the NNLO contributions, at $\mathcal{O}(\alpha_S^6)$ in the case of double $J/\psi$ hadroproduction. The cancellation of the aforementioned infrared divergences would be achieved as usual by considering two-loop, one-loop single-real-emission and double-real-emission amplitudes. Such a computation is obviously  beyond the scope of this study -- it is not even available for single $J/\psi$.

However, a subset of such one-loop diagrams, restricted to the sole topologies with two separate charm-quark lines forming each a $J/\psi$, happens to be free of any divergence and is, in addition, gauge invariant. Correspondingly, the possible  double-real emissions which could develop infrared divergences do not contribute when one of the external gluon becomes soft. This is akin to the absence of any infrared divergences at $P_T \to 0$ for $gg \to J/\psi g$. Such a subset is in fact that of the LI contribution to $pp\rightarrow J/\psi+\Upsilon$~considered in Ref.~\cite{Shao:2016wor}

The square of the amplitude from these one-loop diagrams is what we refer here to as the (partial) LI corrections. Their computation is included in the {\sc\small HELAC-Onia} code~\cite{Shao:2012iz,Shao:2015vga} and is thus available to everybody. In fact, another gauge-invariant  $\mathcal{O}(\alpha_S^6)$ part, namely from $pp\rightarrow J/\psi+J/\psi+c\bar{c}$, is known~\cite{Lansberg:2014swa}. It turns out to be small and can safely be ignored for our purposes. However, we wish to point out that the process $pp\rightarrow J/\psi+J/\psi+c\bar{c}$ has its own interest as it can be a potential probe of the TPS at the LHC.

Let us add that we do not expect any specific kinematical enhancement of other NNLO topologies, in particular that of the double-real-gluon emission in view of the results of $pp\rightarrow J/\psi+\Upsilon$~\cite{Shao:2016wor}. This is partly explained by the vanishing of these contributions when one gluon becomes soft, precisely where one can minimise the off-shellness of the other particles involved in the scattering and thus where these contributions could have been the largest.

\begin{figure}[hbt!]
\centering
\foreach \page in {3,1}{
\includegraphics[page=\numexpr \page\relax, width=.48\textwidth,draft=false]{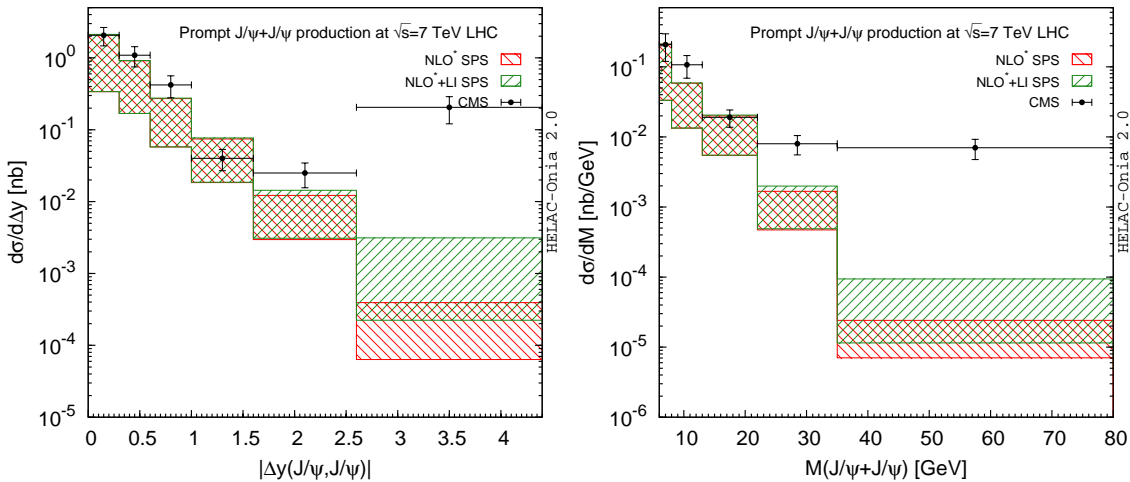}
}
\caption{Rapidity gap $|\Delta y(J/\psi, J/\psi)|$ (left) and invariant mass $M(J/\psi+J/\psi)$ (right) distributions for di-$J/\psi$ production in CSM via SPS within CMS $\sqrt{s}=7$ TeV acceptance~\cite{Khachatryan:2014iia}.}\label{fig:cmsli}
\end{figure}

\begin{figure}[hbt!]
\centering
\foreach \page in {4,1}{
\includegraphics[page=\numexpr \page\relax, width=.48\textwidth,draft=false]{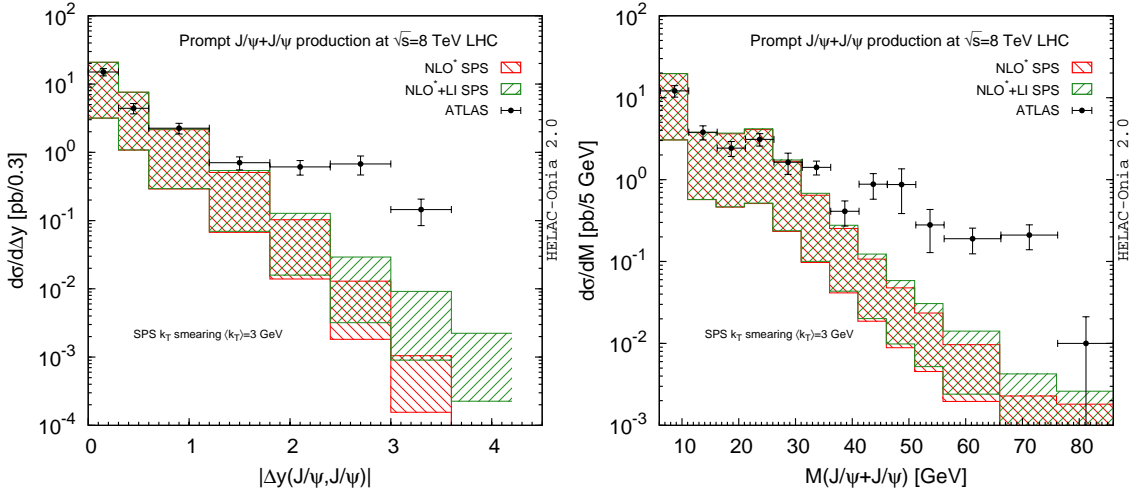}
}
\caption{Rapidity gap $|\Delta y(J/\psi, J/\psi)|$ (left) and invariant mass $M(J/\psi+J/\psi)$ (right) distributions for di-$J/\psi$ production in CSM via SPS within ATLAS $\sqrt{s}=8$ TeV acceptance~\cite{Aaboud:2016fzt}.}\label{fig:atlasli}
\end{figure}

\begin{figure}[hbt!]
\centering
\foreach \page in {8,2}{
\includegraphics[page=\numexpr \page\relax, width=.48\textwidth,draft=false]{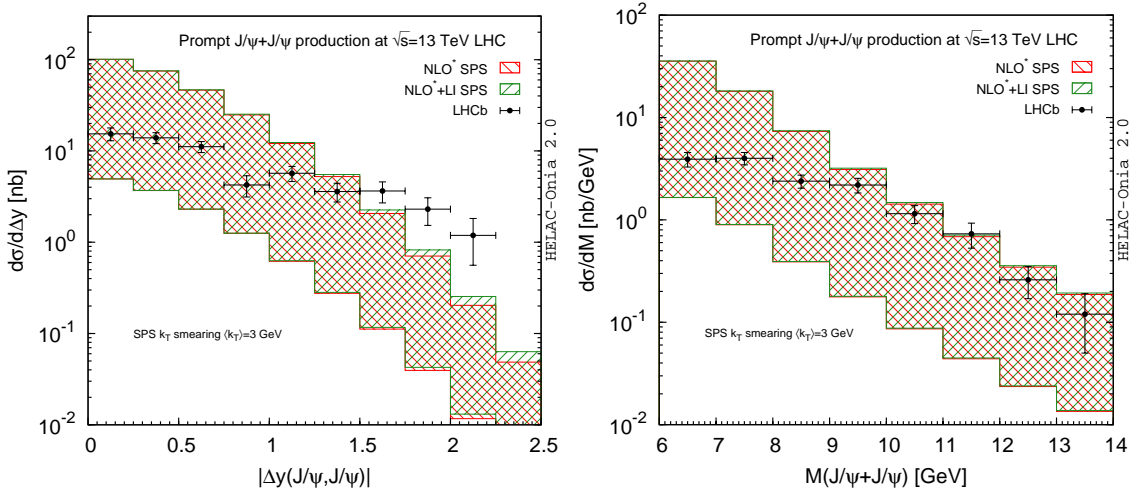}
}
\caption{Rapidity gap $|\Delta y(J/\psi, J/\psi)|$ (left) and invariant mass $M(J/\psi+J/\psi)$ (right) distributions for di-$J/\psi$ production in CSM via SPS within LHCb $\sqrt{s}=13$ TeV acceptance~\cite{Aaij:2016bqq}.}\label{fig:lhcbli}
\end{figure}

The cross-sections differential in the absolute rapidity difference between the $J/\psi$ pair $|\Delta y(J/\psi,J/\psi)|$ are shown in the left panels of Figures~\ref{fig:cmsli},~\ref{fig:atlasli} and \ref{fig:lhcbli} and are compared to the CMS, ATLAS and LHCb data. The NLO$^\star$ CS calculations are displayed by the red hatched bands in the figures. The partial LI contributions are represented by the green bands. As expected, the (partial) LI is significant at large $|\Delta y(J/\psi,J/\psi)|$ region but negligible at small and intermediate $|\Delta y|$. An order of magnitude enhancement to the CS cross section is expected when $|\Delta y(J/\psi,J/\psi)|\geq 3.0$. Nonetheless, despite the very large theoretical uncertainties from the scale variations, a discrepancy between the CS SPS and the experimental data is clearly visible at large $|\Delta y(J/\psi,J/\psi)|$, that is exactly where the DPS is expected to be important. 

The invariant mass of the meson pair is also closely related to the rapidity gap $|\Delta y(J/\psi,J/\psi)|$ (see the discussion in Ref.~\cite{Lansberg:2014swa}). Large $M(J/\psi+J/\psi)$ bins are usually populated by large $|\Delta y(J/\psi,J/\psi)|$ events. Similar enhancements from the LI contributions can be seen in the tail of the invariant-mass distributions of Figures~\ref{fig:cmsli} and \ref{fig:atlasli}. The measurements by CMS and ATLAS are consistent with the SPS CS alone at low invariant masses and depart from the SPS CS bands (NLO$^\star$ and NLO$^\star$+LI) at  large $M(J/\psi+J/\psi)$ values. In contrast, for the LHCb acceptance, the LI part is negligible compared to the NLO$^\star$ contributions due to the limited  $M(J/\psi+J/\psi)$ range, below 14 GeV. 

We have collected additional data-theory-comparison plots between the SPS CS yields and the LHC measurements for other observables in the appendix \ref{app:LImoreplots}. The data are compatible with the CS theoretical predictions but the LI contributions are found to be negligible for all the other distributions.




\section{Comprehensive assessment of the colour-octet contributions\label{sec:copart}}

The whole LO CO contributions to di-$\psi$ hadroproduction at the LHC up to $\mathcal{O}(v^7)$ in NRQCD have recently been computed~by He and Kniehl~\cite{He:2015qya}. Their study however bears on a single CO LDME set from an out-of-date LO single $J/\psi$ hadroproduction fit~\cite{Braaten:1999qk} which was made with the early Tevatron data. Yet, their calculation seems to indicate that the CO contributions might be relevant at large $|\Delta y(J/\psi,J/\psi)|$ and large $M(J/\psi+J/\psi)$ due to similar $t$-channel gluon exchange diagrams than for the CS LI contributions. The aforementioned remaining discrepancy between this full SPS LO NRQCD calculation and the CMS data at large $|\Delta y(J/\psi,J/\psi)|$ was then attributed to unknown missing higher-order QCD corrections to the CO contributions.

We however note that we do not anticipate any such so-called ``giant" $K$ factors in this region. Currently, no complete NLO CO calculation exists. Since it is important to deal with a complete set of CO channels in order to guarantee the large cancellation between $S$-wave and $P$-wave contributions involved the NLO LDME fits of hadroproduction data, we consider that to rely on a LO --but complete-- perturbative calculation and then to estimate the size of the missing higher-order corrections via the scale uncertainty is probably the most reasonable procedure to adopt.

An alternative approach to investigate the presence of possible ``giant" $K$ factors from new fragmentation topologies --if some are indeed relevant-- without performing a full computation is that recently proposed by one of us in Ref.~\cite{Shao:2018adj}. It has been proved useful for the single $J/\psi$ hadroproduction case. The method is in principle general and applicable for the double $J/\psi$ hadroproduction as well, although a new infrared divergence in double $P$-wave channels emerges~\cite{He:2018hwb}. We leave it for future studies since it may not apply to the whole phase space which we wish to consider here. Finally, we note that a similar enhancement from $t$-channel gluon exchange was expected for di-$\chi_c$ production but its feed down was also found to be insignificant in the di-$\psi$ yield~\cite{Cisek:2017ikn}.

\subsection{Status and issues with the colour-octet transitions}

Although the possibility for CO transitions is a robust prediction from NRQCD, their actual impact in the phenomenology has been the subject of debates for decades. The most glaring observations for the necessity of their presence are twofold. First, CO provide a natural solution for the infra-red divergence issue in $P$-wave production. Second, the LO $v^2$ NRQCD calculation involving only CS transitions still underestimates --even after including NLO QCD corrections-- the yields of single $J/\psi$ and $\psi(2S)$ hadroproduction at large $P_T$ at the Tevatron and the LHC. 

However, NRQCD computations even including CO contributions are unable to coherently describe --{\it i.e.} with the same CO LDMEs-- the world data for $pp$, $ep$, $\gamma p$, $\gamma \gamma$ and $e^+e^-$ collisions. For a recent review, we guide the reader to Ref.~\cite{Lansberg:2019adr}.

The CO LDMEs are predicted to be universal non-perturbative objects by NRQCD, which should yield predictions compatible with all the data. The current status of their extractions is very confusing as their numerical values and their uncertainties are very disparate. The results of the fits of different groups disagree with each others. As long as the situation is not clarified, we believe that it is necessary to comprehensively consider these analyses instead of drawing conclusions based on a single CO LDME set as it is often done in the analysis of associated production of quarkonium (see Ref.~\cite{Lansberg:2019adr} for some examples).

As such, we will use different LDME sets of which we briefly review the status and the possible limitations. As we said above, the available CO LDMEs for prompt $J/\psi$ production are extracted from fits. According to the QCD accuracy of the short-distance coefficients (SDCs), we will categorise them in the 4 groups shown in Table~\ref{tab:coldme}. Namely, 
\begin{enumerate}
\item three fits are based on LO SDCs~\cite{Braaten:1999qk,Kramer:2001hh,Sharma:2012dy}, 
\item four fits based on NLO SDCs~\cite{Butenschoen:2011yh,Gong:2012ug,Shao:2014yta,Han:2014jya},  
\item one fit based on a low-$P_T$ leading-logarithm (LL) resummed SDC ~\cite{Sun:2012vc},
\item one fit using a SDC using leading-power (LP) fragmentation matched to NLO SDC~\cite{Bodwin:2014gia}.
\end{enumerate} 

All of them have shortcomings and/or limitations. We enumerate them below:
\begin{enumerate}
\item First of all, we wish to emphasise that the LO fits are out-of-date and should be viewed as a pure tuning of the normalisation of the single $J/\psi$ data. Since all of the LO fits are mainly performed with the help of intermediate and large $P_T$ hadroproduction data, where the ``giant" $K$ factors from NLO QCD corrections emerge, it is very hard to imagine that these values will give correct predictions for independent observables, like the double-$J/\psi$ hadroproduction in our case, for which $K$ factors would be different. We will therefore use them here for a pure illustrative purpose.
\item The LL fit in Ref.~\cite{Sun:2012vc} concentrates on the $P_T(J/\psi)< m_c$ region. The authors performed a small-$P_T$ resummation but without considering the contribution from the CS channel which is however known to saturate the data in this region~\cite{Brodsky:2009cf,Feng:2015cba}. The values of these LDMEs have never been used for the single $J/\psi$ production at intermediate and large $P_T$ regions. They are included in our discussion like the LO fits in order to be exhaustive.
\item The NLO fit in Ref.~\cite{Butenschoen:2011yh} used the world data before 2011 without subtracting the feed-down contributions. The fit seems to yield a good agreement with the $P_T(J/\psi)<30$ GeV $J/\psi$ yields data at different colliders but for $\gamma \gamma$ and $e^+e^-$ collisions. However, it overshoots the $P_T>30$ GeV yields and fails to reproduce the polarisations of $J/\psi$, the energy-fraction distribution of the $J/\psi$ in jets~\cite{Bain:2017wvk} and the yields of $\eta_c$ (by using heavy-quark spin symmetry). In addition, the SPS $P_T$-differential cross section of $J/\psi+\gamma$~\cite{Li:2014ava} turns out to be negative at NLO with these values of CO LDMEs.
\item The NLO fit by Gong et al.~\cite{Gong:2012ug} focus on the $P_T(J/\psi)>7$ GeV data. The feed-down contributions are subtracted. This LDME set is however not compatible with the yields (\eg\ $pp$, $\gamma p$ and $e^+e^-$) when $P_T(J/\psi)<7$ GeV, the polarisation of forward $J/\psi$~\cite{Aaij:2013nlm} and the $\eta_c$ production. In addition, it yields to  --unphysical-- negative cross sections in $pp\rightarrow J/\psi+\gamma$. In principle, this set is only applicable to $J/\psi$ production with $P_T(J/\psi)>7$ GeV, \ie\ only to the ATLAS fiducial region for our forthcoming discussion of double $J/\psi$ production.
\item The two sets denoted sets 7 and 8 in Table~\ref{tab:coldme} are two extreme cases of the PKU fit~\cite{Shao:2014yta,Han:2014jya} after including the constraints from LHC $\eta_c$ data~\cite{Shao:2014yta,Han:2014jya}. They  supersede the fits including the $P_T(J/\psi)>7$ GeV hadroproduction data described in Refs.~\cite{Ma:2010yw,Chao:2012iv}. These LDME sets cannot reproduce the CDF polarisation measurement~\cite{Abulencia:2007us} --like all the other sets in fact-- and are not applicable to $P_T(J/\psi)<7$ GeV. Both sets should only be used to di-$\psi$ production in the ATLAS fiducial region.
\item The NLO+LP fit of Ref.~\cite{Bodwin:2014gia} --as well as its update~\cite{Bodwin:2015iua}-- has been presented by its authors as the only fit able to reproduce the $J/\psi$ data (both yields and polarisations) above $10$ GeV after including the LP fragmentation contributions on top of the NLO calculations. However, it does not yield the correct $\eta_c$ cross section in the same $P_T$ region under the heavy-quark spin symmetry. As what concerns predictions for double $J/\psi$ production, it is marginally applicable only in the ATLAS fiducial region with $P_T(J/\psi)>8.5$ GeV instead of 10 GeV.
\end{enumerate}

Since we aim at a comprehensive analysis, we have considered all of the $9$ sets listed in Table~\ref{tab:coldme} to show how strongly the CO contributions depend on the available CO LDMEs. We should however recall during the discussion what we believe to be the region of applicability in $P_T(J/\psi)$ for the NLO(+LP) fits.

\begin{table*}[htpb]
\begin{center} \small

\begin{tabular}{c|c|c|c|c|c}
\cline{1-5}
& \multicolumn{3}{c}{LO fits} & \multicolumn{1}{|c|}{LL fit} & \\\cline{1-5}
& Set 1 & Set 2 & Set 3 & Set 4 & \\
\cline{1-5}
$\langle \mathcal{O}^{J/\psi}(^3S_1^{[1]})\rangle$ [GeV$^3$] & $1.2$ GeV$^3$ & $1.4$  & $1.16$  & $1.16$  & \\
$\langle \mathcal{O}^{J/\psi}(^3S_1^{[8]})\rangle$ [GeV$^3$]& $1.3\cdot 10^{-3}$  & $3.9\cdot 10^{-3}$ & $1.2\cdot 10^{-2}$ & $-9.3\cdot 10^{-3}$ & \\
$\langle \mathcal{O}^{J/\psi}(^1S_0^{[8]})\rangle$ [GeV$^3$]& $1.8\cdot 10^{-2}$  & $0$ & $0$ & $0.14$  & \\
$\langle \mathcal{O}^{J/\psi}(^3P_0^{[8]})\rangle$ [GeV$^5$]& $3.5\cdot 10^{-2}$ & $4.4\cdot 10^{-2}$  & $2.9\cdot 10^{-2}$  & $-3.9\cdot 10^{-2}$  & \\
$\langle \mathcal{O}^{\psi(2S)}(^3S_1^{[1]})\rangle$ [GeV$^3$]& $0.76$ & $0.67$  & $0.76$  & $0$ & \\
$\langle \mathcal{O}^{\psi(2S)}(^3S_1^{[8]})\rangle$ [GeV$^3$]& $3.3\cdot 10^{-3}$  & $3.7\cdot 10^{-3}$ & $5\cdot 10^{-3}$ & $0$ & \\
$\langle \mathcal{O}^{\psi(2S)}(^1S_0^{[8]})\rangle$ [GeV$^3$]& $8.0\cdot 10^{-3}$ & $0$ & $0$ & $0$ & \\
$\langle \mathcal{O}^{\psi(2S)}(^3P_0^{[8]})\rangle$ [GeV$^5$]& $1.6\cdot 10^{-2}$  & $5.0\cdot 10^{-3}$  & $1.2\cdot 10^{-2}$  & $0$ & \\
$\langle \mathcal{O}^{\chi_{c0}}(^3S_1^{[8]})\rangle$ [GeV$^3$]& $1.9\cdot 10^{-3}$  & $1.9\cdot 10^{-3}$ & $3.1\cdot 10^{-3}$  & $0$ & \\
$\langle \mathcal{O}^{\chi_{c0}}(^3P_0^{[1]})\rangle$ [GeV$^5$]& $0.11$  & $9.1\cdot 10^{-2}$  & $0.11$ & $0$ & \\
\hline
& \multicolumn{4}{c}{NLO fits} & \multicolumn{1}{|c}{NLO+LP fit} \\\hline
& Set 5 & Set 6 & Set 7 & Set 8 & Set 9 \\
\hline
$\langle \mathcal{O}^{J/\psi}(^3S_1^{[1]})\rangle$ [GeV$^3$]& $1.32$ GeV$^3$ & $1.16$ & $1.16$  & $1.16$  & $1.16$  \\
$\langle \mathcal{O}^{J/\psi}(^3S_1^{[8]})\rangle$ [GeV$^3$]& $2.2\cdot 10^{-3}$  & $-4.6\cdot 10^{-3}$  &  $1.1\cdot 10^{-2}$  & $9.0\cdot 10^{-3}$  & $1.1\cdot 10^{-2}$  \\
$\langle \mathcal{O}^{J/\psi}(^1S_0^{[8]})\rangle$ [GeV$^3$]& $5.0\cdot 10^{-2}$  & $9.7\cdot 10^{-2}$  & $0$ & $1.5\cdot 10^{-2}$ & $9.9\cdot 10^{-2}$  \\
$\langle \mathcal{O}^{J/\psi}(^3P_0^{[8]})\rangle$ [GeV$^5$]& $-1.6\cdot 10^{-2}$ & $-2.1\cdot 10^{-2}$& $4.2\cdot 10^{-2}$  & $3.4\cdot 10^{-2}$ & $1.1\cdot 10^{-2}$ \\
$\langle \mathcal{O}^{\psi(2S)}(^3S_1^{[1]})\rangle$ [GeV$^3$]& $0$ & $0.76$  & $0.76$  & $0.76$ & $0$ \\
$\langle \mathcal{O}^{\psi(2S)}(^3S_1^{[8]})\rangle$ [GeV$^3$]& $0$ & $3.4\cdot 10^{-3}$  & $6.1\cdot 10^{-3}$  & $1.2\cdot 10^{-3}$  & $0$ \\
$\langle \mathcal{O}^{\psi(2S)}(^1S_0^{[8]})\rangle$ [GeV$^3$]& $0$ & $-1.2\cdot 10^{-4}$  & $0$ & $2.0\cdot 10^{-2}$  & $0$ \\
$\langle \mathcal{O}^{\psi(2S)}(^3P_0^{[8]})\rangle$ [GeV$^5$]& $0$ & $9.5\cdot 10^{-3}$  & $2.2\cdot 10^{-2}$  & $0$ & $0$\\
$\langle \mathcal{O}^{\chi_{c0}}(^3S_1^{[8]})\rangle$ [GeV$^3$]& $0$ & $2.2\cdot 10^{-3}$  & $2.2\cdot 10^{-3}$  & $2.2\cdot 10^{-3}$  & $0$\\
$\langle \mathcal{O}^{\chi_{c0}}(^3P_0^{[1]})\rangle$ [GeV$^5$]& $0$ & $0.11$ & $0.11$  & $0.11$ & $0$\\\hline
$P_T(J/\psi)$ region & $<30$ GeV & $>7$ GeV & $>7$ GeV & $>7$ GeV & $>10$ GeV \\
\hline
\end{tabular}
\caption{Summary of LDMEs we used from various fits [Set 1: Sharma et al.~\cite{Sharma:2012dy}; Set 2: Braaten et al.~\cite{Braaten:1999qk}; Set 3: Kr{\"a}mer~\cite{Kramer:2001hh};  Set 4: Sun et al.~\cite{Sun:2012vc}; Set 5: Butensch{\"o}n et al.~\cite{Butenschoen:2011yh}; Set 6 : Gong et al.~\cite{Gong:2012ug}; Set 7: Shao et al.~\cite{Shao:2014yta}: Set 8: Han et al.~\cite{Han:2014jya}: Set 9: Bodwin et al.~\cite{Bodwin:2014gia}].}
\label{tab:coldme}
\end{center}
\end{table*}

\subsection{Numerical results}

In this section, we will present our numerical results with the LO CO channels summed to the pure NLO$^\star$ CS channel $^3S_1^{[1]}+^3S_1^{[1]}$. Although the CS LDMEs $\langle\mathcal{O}^{J/\psi}(^3S_1^{[1]})\rangle$ and $\langle\mathcal{O}^{\psi(2S)}(^3S_1^{[1]})\rangle$ vary from set to set in Table \ref{tab:coldme}, we will fix these values for the NLO$^\star$ CS channel $^3S_1^{[1]}+^3S_1^{[1]}$ to those used in section ~\ref{sec:setup}. The uncertainty from these LDMEs is systematically subdominant compared to the scale uncertainty. All the feed-down contributions are properly taken into account as well.

\subsubsection{LHCb data at $\sqrt{s}=13$ TeV}

We start our discussion with the LHCb acceptance~\cite{Aaij:2016bqq}, where the $P_T(J/\psi)$ can be as low as zero. We have compared the CSM NLO$^\star$+COM LO SPS (the green bands) with the data in Figure~\ref{fig:colhcbdy} for the $\Delta y(J/\psi,J/\psi)$ distribution and in Figure~\ref{fig:colhcbdm} for the invariant mass of the pair $M(J/\psi+J/\psi)$ distribution. Like we have found for the CS LI contributions, the CO contributions are not relevant in the invariant mass distribution of LHCb. They start to be slightly visible in the tail of the $\Delta y(J/\psi,J/\psi)$ distribution.

\begin{figure}[hbt!]
\centering

\foreach \page/\ipage in {2/1}{
\subfloat[Set \ipage]{\includegraphics[page=\numexpr \page\relax, scale=.32,draft=false,draft=false,trim = 0mm 0mm 0mm 0mm,clip]{\LHCbPlotDir/dy_sigma_LHCbarXiv161207451_CO.pdf}}
}
\hspace*{-.35cm}
\foreach \page/\ipage in {9/2,8/3}{
\subfloat[Set \ipage]{\includegraphics[page=\numexpr \page\relax, scale=.32,draft=false,trim = 21.8mm 0mm 0mm 0mm,clip]{\LHCbPlotDir/dy_sigma_LHCbarXiv161207451_CO.pdf}}\hspace*{-.2cm}
}

\foreach \page/\ipage in {4/4}{
\subfloat[Set \ipage]{\includegraphics[page=\numexpr \page\relax, scale=.32,draft=false,trim = 0mm 0mm 0mm 0mm,clip]{\LHCbPlotDir/dy_sigma_LHCbarXiv161207451_CO.pdf}}
}
\hspace*{-.35cm}
\foreach \page/\ipage in {1/5,3/6}{
\subfloat[Set \ipage]{\includegraphics[page=\numexpr \page\relax, scale=.32,draft=false,trim = 21.8mm 0mm 0mm 0mm,clip]{\LHCbPlotDir/dy_sigma_LHCbarXiv161207451_CO.pdf}}\hspace*{-.2cm}
}

\foreach \page/\ipage in {6/7}{
\subfloat[Set \ipage]{\includegraphics[page=\numexpr \page\relax, scale=.32,draft=false]{\LHCbPlotDir/dy_sigma_LHCbarXiv161207451_CO.pdf}}
}
\hspace*{-.35cm}
\foreach \page/\ipage in {7/8,5/9}{
\subfloat[Set \ipage]{\includegraphics[page=\numexpr \page\relax, scale=.32,draft=false,trim = 21.8mm 0mm 0mm 0mm,clip]{\LHCbPlotDir/dy_sigma_LHCbarXiv161207451_CO.pdf}\hspace*{-.2cm}}
}



\caption{$\Delta y(J/\psi,J/\psi)$ distributions in NLO$^\star$ CS and LO CO via SPS within LHCb $\sqrt{s}=13$ TeV acceptance~\cite{Aaij:2016bqq}.}\label{fig:colhcbdy}
\end{figure}

\begin{figure}[hbt!]
\centering
\foreach \page/\ipage in {2/1}{
\subfloat[Set \ipage]{\includegraphics[page=\numexpr \page\relax,scale=.32,draft=false]{\LHCbPlotDir/dM_sigma_LHCbarXiv161207451_CO.pdf}}
}
\hspace*{-.35cm}
\foreach \page/\ipage in {9/2,8/3}{
\subfloat[Set \ipage]{\includegraphics[page=\numexpr \page\relax,scale=.32,draft=false,trim = 21.8mm 0mm 0mm 0mm,clip]{\LHCbPlotDir/dM_sigma_LHCbarXiv161207451_CO.pdf}\hspace*{-.2cm}}
}

\foreach \page/\ipage in {4/4}{
\subfloat[Set \ipage]{\includegraphics[page=\numexpr \page\relax,scale=.32,draft=false]{\LHCbPlotDir/dM_sigma_LHCbarXiv161207451_CO.pdf}}
}
\hspace*{-.35cm}
\foreach \page/\ipage in {1/5,3/6}{
\subfloat[Set \ipage]{\includegraphics[page=\numexpr \page\relax,scale=.32,draft=false,trim = 21.8mm 0mm 0mm 0mm,clip]{\LHCbPlotDir/dM_sigma_LHCbarXiv161207451_CO.pdf}\hspace*{-.2cm}}
}

\foreach \page/\ipage in {6/7}{
\subfloat[Set \ipage]{\includegraphics[page=\numexpr \page\relax,scale=.32,draft=false]{\LHCbPlotDir/dM_sigma_LHCbarXiv161207451_CO.pdf}}
}
\hspace*{-.35cm}
\foreach \page/\ipage in {7/8,5/9}{
\subfloat[Set \ipage]{\includegraphics[page=\numexpr \page\relax,scale=.32,draft=false,trim = 21.8mm 0mm 0mm 0mm,clip]{\LHCbPlotDir/dM_sigma_LHCbarXiv161207451_CO.pdf}\hspace*{-.2cm}}
}

\caption{$M(J/\psi+J/\psi)$ distributions in NLO$^\star$ CS and LO CO via SPS within LHCb $\sqrt{s}=13$ TeV acceptance~\cite{Aaij:2016bqq}.}\label{fig:colhcbdm}
\end{figure}

This observation however very much depends on the set of CO LDMEs used. In particular, the only plausible set, \ie\ set 5, in the small $P_T(J/\psi)$ region does not yield any significant contribution to the cross section. It also seems clear that none of the sets can fully account for the discrepancy between SPS and LHCb data in the last bins of $\frac{d\sigma}{d\Delta y}$. Additional plots for the comparisons between CS NLO$^\star$+CO LO SPS and data can be found in appendix~\ref{app:comoreplots}. The impact of the CO contributions on these additional distributions is in general minor.

\subsubsection{CMS data at $\sqrt{s}=7$ TeV}

The events analysed by CMS have larger $P_T(J/\psi)$ above $4.5$ GeV to $6.5$ GeV depending on the rapidity. In this region, the only applicable NLO fit is still set 5 taken from Ref.~\cite{Butenschoen:2011yh}. As opposed to the conclusion made in Ref.~\cite{He:2015qya}, the CO SPS contribution is either much suppressed compared to the CS SPS contributions or much smaller than the experimental data as shown in Figure~\ref{fig:cocmsdy} and Figure~\ref{fig:cocmsdm}. Given that the LO fits (like that used in Ref.~\cite{He:2015qya} (\ie\ set 2))  are not plausible any more and that the only applicable fit is the NLO fit given by set 5, we draw the conclusion that our extraction of DPS in Ref.~\cite{Lansberg:2014swa} --made by neglecting the CO contributions-- is still  sound, which actually has been shown to be consistent with the ATLAS measurement thanks to a completely different method to disentangle the DPS from the SPS contributions and in a different kinematical region.

\begin{figure}[hbt!]
\centering
\foreach \page/\ipage in {2/1}{
\subfloat[Set \ipage]{\includegraphics[page=\numexpr \page\relax,scale=.32,draft=false]{\CMSPlotDir/dy_sigma_CMSarXiv14060484_CO.pdf}}
}
\hspace*{-.35cm}
\foreach \page/\ipage in {9/2,8/3}{
\subfloat[Set \ipage]{\includegraphics[page=\numexpr \page\relax,scale=.32,trim = 21.8mm 0mm 0mm 0mm,clip,draft=false]{\CMSPlotDir/dy_sigma_CMSarXiv14060484_CO.pdf}\hspace*{-.2cm}}
}

\foreach \page/\ipage in {4/4}{
\subfloat[Set \ipage]{\includegraphics[page=\numexpr \page\relax,scale=.32,draft=false]{\CMSPlotDir/dy_sigma_CMSarXiv14060484_CO.pdf}}
}
\hspace*{-.35cm}
\foreach \page/\ipage in {1/5,3/6}{
\subfloat[Set \ipage]{\includegraphics[page=\numexpr \page\relax,scale=.32,trim = 21.8mm 0mm 0mm 0mm,clip,draft=false]{\CMSPlotDir/dy_sigma_CMSarXiv14060484_CO.pdf}\hspace*{-.2cm}}
}

\foreach \page/\ipage in {6/7}{
\subfloat[Set \ipage]{\includegraphics[page=\numexpr \page\relax,scale=.32,draft=false]{\CMSPlotDir/dy_sigma_CMSarXiv14060484_CO.pdf}}
}
\hspace*{-.35cm}
\foreach \page/\ipage in {7/8,5/9}{
\subfloat[Set \ipage]{\includegraphics[page=\numexpr \page\relax,scale=.32,trim = 21.8mm 0mm 0mm 0mm,clip,draft=false]{\CMSPlotDir/dy_sigma_CMSarXiv14060484_CO.pdf}\hspace*{-.2cm}}
}

\caption{$\Delta y(J/\psi,J/\psi)$ distributions in NLO$^\star$ CS and LO CO via SPS within CMS $\sqrt{s}=7$ TeV acceptance~\cite{Khachatryan:2014iia}.}\label{fig:cocmsdy}
\end{figure}

\begin{figure}[hbt!]
\centering
\foreach \page/\ipage in {2/1}{
\subfloat[Set \ipage]{\includegraphics[page=\numexpr \page\relax,scale=.32,draft=false]{\CMSPlotDir/dM_sigma_CMSarXiv14060484_CO.pdf}}
}
\hspace*{-.35cm}
\foreach \page/\ipage in {9/2,8/3}{
\subfloat[Set \ipage]{\includegraphics[page=\numexpr \page\relax,scale=.32,draft=false,trim = 21.8mm 0mm 0mm 0mm,clip]{\CMSPlotDir/dM_sigma_CMSarXiv14060484_CO.pdf}\hspace*{-.2cm}}
}

\foreach \page/\ipage in {4/4}{
\subfloat[Set \ipage]{\includegraphics[page=\numexpr \page\relax,scale=.32,draft=false]{\CMSPlotDir/dM_sigma_CMSarXiv14060484_CO.pdf}}
}
\hspace*{-.35cm}
\foreach \page/\ipage in {1/5,3/6}{
\subfloat[Set \ipage]{\includegraphics[page=\numexpr \page\relax,scale=.32,draft=false,trim = 21.8mm 0mm 0mm 0mm,clip]{\CMSPlotDir/dM_sigma_CMSarXiv14060484_CO.pdf}\hspace*{-.2cm}}
}

\foreach \page/\ipage in {6/7}{
\subfloat[Set \ipage]{\includegraphics[page=\numexpr \page\relax,scale=.32,draft=false]{\CMSPlotDir/dM_sigma_CMSarXiv14060484_CO.pdf}}
}
\hspace*{-.35cm}
\foreach \page/\ipage in {7/8,5/9}{
\subfloat[Set \ipage]{\includegraphics[page=\numexpr \page\relax,scale=.32,draft=false,trim = 21.8mm 0mm 0mm 0mm,clip]{\CMSPlotDir/dM_sigma_CMSarXiv14060484_CO.pdf}\hspace*{-.2cm}}
}

\caption{$M(J/\psi,J/\psi)$ distributions in NLO$^\star$ CS and LO CO via SPS within CMS $\sqrt{s}=7$ TeV acceptance~\cite{Khachatryan:2014iia}.}\label{fig:cocmsdm}
\end{figure}




\subsubsection{ATLAS data at $\sqrt{s}=8$ TeV}

\begin{figure}[hbt!]
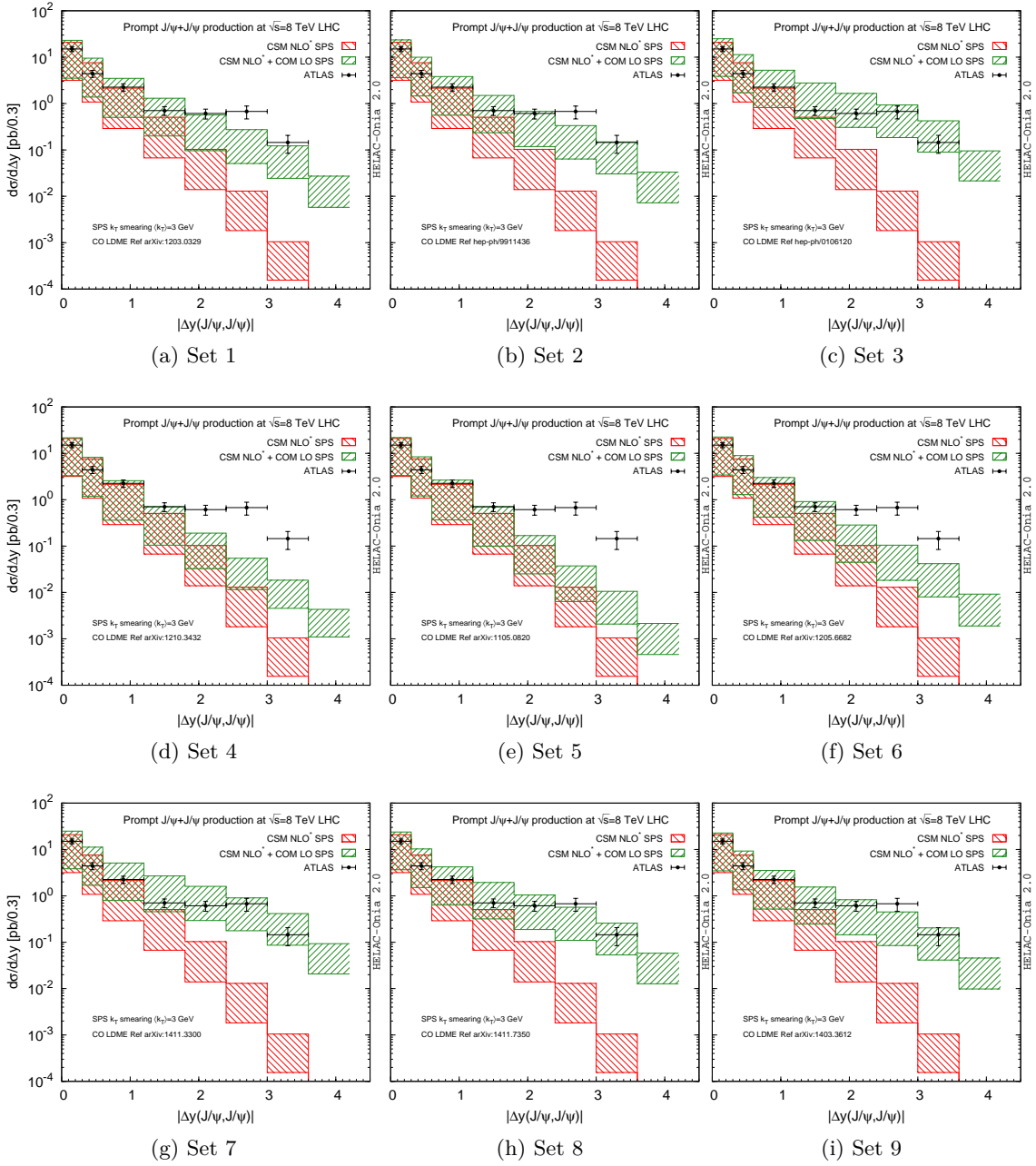

\centering
\foreach \page/\ipage in {2/1}{
\subfloat[Set \ipage]{\includegraphics[page=\numexpr \page\relax,scale=.32,draft=false]{\ATLASPlotDir/dy_sigma_ATLASCONF2016047_CO.pdf}}
}
\hspace*{-.35cm}
\foreach \page/\ipage in {9/2,8/3}{
\subfloat[Set \ipage]{\includegraphics[page=\numexpr \page\relax,scale=.32,draft=false,trim = 21.8mm 0mm 0mm 0mm,clip]{\ATLASPlotDir/dy_sigma_ATLASCONF2016047_CO.pdf}\hspace*{-.2cm}}
}

\foreach \page/\ipage in {4/4}{
\subfloat[Set \ipage]{\includegraphics[page=\numexpr \page\relax,scale=.32,draft=false]{\ATLASPlotDir/dy_sigma_ATLASCONF2016047_CO.pdf}}
}
\hspace*{-.35cm}
\foreach \page/\ipage in {1/5,3/6}{
\subfloat[Set \ipage]{\includegraphics[page=\numexpr \page\relax,scale=.32,draft=false,trim = 21.8mm 0mm 0mm 0mm,clip]{\ATLASPlotDir/dy_sigma_ATLASCONF2016047_CO.pdf}\hspace*{-.2cm}}
}

\foreach \page/\ipage in {6/7}{
\subfloat[Set \ipage]{\includegraphics[page=\numexpr \page\relax,scale=.32,draft=false]{\ATLASPlotDir/dy_sigma_ATLASCONF2016047_CO.pdf}}
}
\hspace*{-.35cm}
\foreach \page/\ipage in {7/8,5/9}{
\subfloat[Set \ipage]{\includegraphics[page=\numexpr \page\relax,scale=.32,draft=false,trim = 21.8mm 0mm 0mm 0mm,clip]{\ATLASPlotDir/dy_sigma_ATLASCONF2016047_CO.pdf}\hspace*{-.2cm}}
}

\caption{$\Delta y(J/\psi,J/\psi)$ distributions in NLO$^\star$ CS and LO CO via SPS within ATLAS $\sqrt{s}=8$ TeV acceptance~\cite{Aaboud:2016fzt}.}\label{fig:coatlasdy}
\end{figure}

\begin{figure}[hbt!]
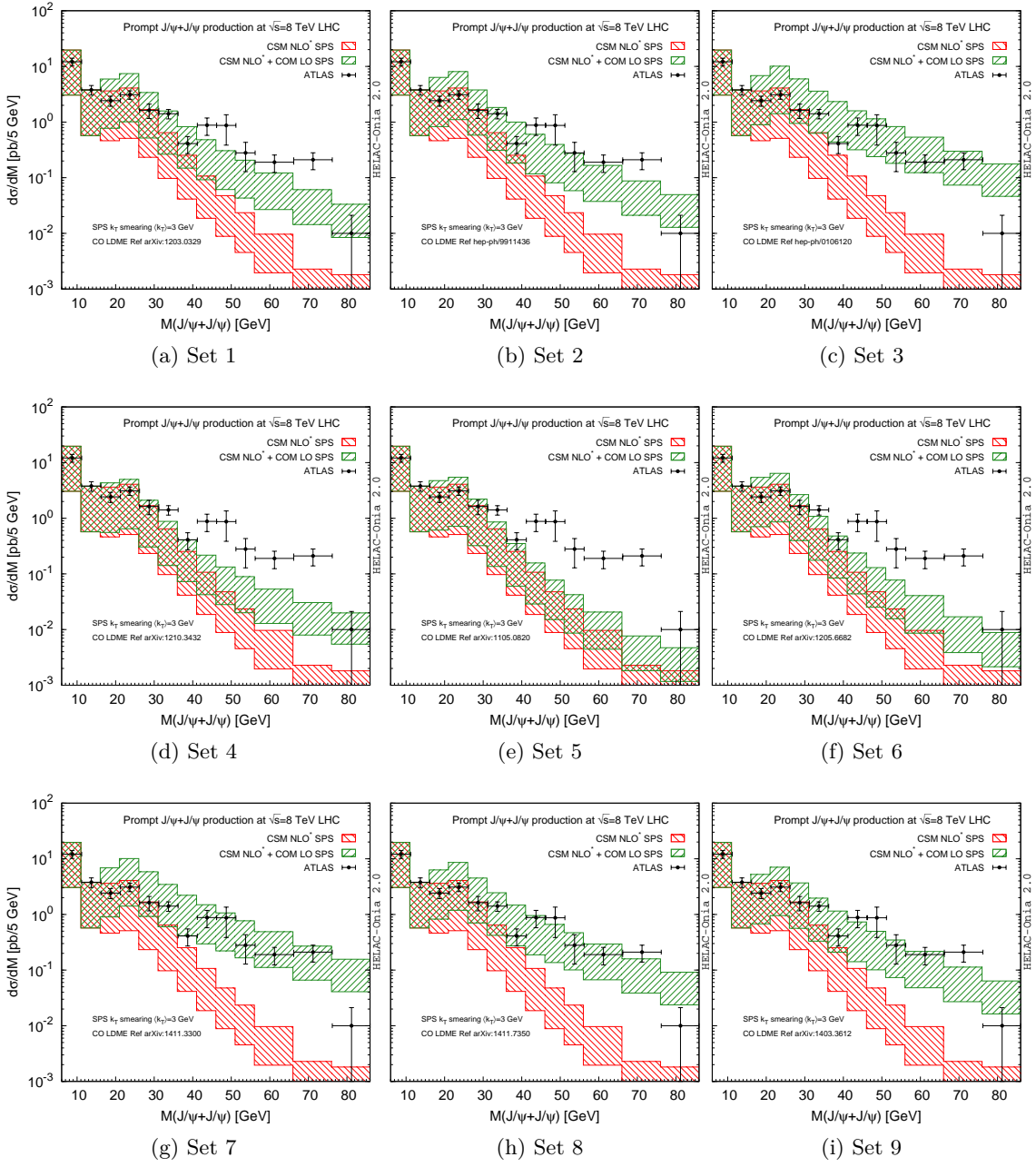

\centering
\foreach \page/\ipage in {2/1}{
\subfloat[Set \ipage]{\includegraphics[page=\numexpr \page\relax,scale=.32,,draft=false]{\ATLASPlotDir/dM_sigma_ATLASCONF2016047_CO.pdf}}}
\hspace*{-.35cm}
\foreach \page/\ipage in {9/2,8/3}{
\subfloat[Set \ipage]{\includegraphics[page=\numexpr \page\relax,scale=.32,,draft=false,trim = 21.8mm 0mm 0mm 0mm,clip]{\ATLASPlotDir/dM_sigma_ATLASCONF2016047_CO.pdf}\hspace*{-.2cm}}}

\foreach \page/\ipage in {4/4}{
\subfloat[Set \ipage]{\includegraphics[page=\numexpr \page\relax,scale=.32,,draft=false]{\ATLASPlotDir/dM_sigma_ATLASCONF2016047_CO.pdf}}}
\hspace*{-.35cm}
\foreach \page/\ipage in {1/5,3/6}{
\subfloat[Set \ipage]{\includegraphics[page=\numexpr \page\relax,scale=.32,,draft=false,trim = 21.8mm 0mm 0mm 0mm,clip]{\ATLASPlotDir/dM_sigma_ATLASCONF2016047_CO.pdf}\hspace*{-.2cm}}}

\foreach \page/\ipage in {6/7}{
\subfloat[Set \ipage]{\includegraphics[page=\numexpr \page\relax,scale=.32,,draft=false]{\ATLASPlotDir/dM_sigma_ATLASCONF2016047_CO.pdf}}}
\hspace*{-.35cm}
\foreach \page/\ipage in {7/8,5/9}{
\subfloat[Set \ipage]{\includegraphics[page=\numexpr \page\relax,scale=.32,,draft=false,trim = 21.8mm 0mm 0mm 0mm,clip]{\ATLASPlotDir/dM_sigma_ATLASCONF2016047_CO.pdf}\hspace*{-.2cm}}}

\caption{$M(J/\psi,J/\psi)$ distributions in NLO$^\star$ CS and LO CO via SPS within ATLAS $\sqrt{s}=8$ TeV acceptance~\cite{Aaboud:2016fzt}.}\label{fig:coatlasdm}
\end{figure}

The transverse momentum cut on single $J/\psi$ is largest in the ATLAS data sample with selected events satisfying $P_T(J/\psi)>8.5$ GeV. This leaves the LDME sets 5-8 as possible good fits. ATLAS used a 2D $(|\Delta y(J/\psi,J/\psi)|,\Delta \phi(J/\psi,J/\psi))$ data-driven template fit to separate SPS and DPS events~\cite{Aaboud:2016fzt}. The control region used to determine the normalisation of DPS is $(|\Delta y(J/\psi,J/\psi)|\geq 1.8,\Delta \phi(J/\psi,J/\psi)\leq \frac{\pi}{2})$. The requirement of $\Delta \phi(J/\psi,J/\psi)\leq \frac{\pi}{2}$ will significantly reduce the CO fraction at large $|\Delta y(J/\psi,J/\psi)|$. The $t$-channel gluon exchange diagrams mainly make the two $J/\psi$ recoiling against each other. It thus populates the region where $\Delta \phi(J/\psi,J/\psi)\rightarrow \pi$. The simultaneous cuts on $|\Delta y(J/\psi,J/\psi)|$ and $\Delta \phi(J/\psi,J/\psi)$ ensure that the DPS extraction in Ref.~\cite{Aaboud:2016fzt} is reliable but for the low statistics in the control region. From Figure~\ref{fig:coatlasdy} and Figure~\ref{fig:coatlasdm}, one sees that the CO yields predicted with the set 7 \& 8 agree reasonably well with the data at large $M(J/\psi,J/\psi)$ and $|\Delta y(J/\psi,J/\psi)|$ with a slight overestimation in the middle of the corresponding distributions. The sets 5 \& 6 however do not agree with the data.  Strong conclusions about the relevance of CO transitions in these regions would thus probably be premature in the absence of a complete NLO study and the disparate values of the existing CO LDMEs.

\section{Conclusions \label{sec:conclusion}}

We have examined two SPS production mechanisms for di-$J/\psi$ production at the LHC, which can be relevant in the control region used to determine the DPS. These are the partial LI CS contributions at $\mathcal{O}(\alpha_S^6)$ and the LO CO contributions at $\mathcal{O}(\alpha_S^4)$. We have also extensively compared our new SPS calculations with the existing LHC data. Our study indeed shows that the LI corrections can enhance the NLO$^\star$ SPS cross section at large $|\Delta y(J/\psi,J/\psi)|$ and large invariant mass $M(J/\psi+J/\psi)$. However, they are not sufficient to explain the discrepancy between SPS theoretical results and the LHC data in these regions. The inclusion of the DPS in the predictions is still crucial to account for the measurements. 

On the other hand, the relevance of the CO contributions in the SPS yield strongly depends on the considered LDME set, thus with a very low predictive power --given the current status of understanding of the COM. It is in any case confined to the large $|\Delta y(J/\psi,J/\psi)|$ region.  We anyhow conclude that the CO contributions can only be important when compared to the ATLAS data but that the ATLAS DPS extraction via a 2D data-driven fit is very likely free of any bias due to a possibly underestimated CO contribution in their control region. Such a conclusion is backed up by studies~\cite{Yamanaka:2018blj,CEM:forthcoming} made in the colour-evaporation model which offers a complementary framework to study the impact of CO transitions.

\begin{acknowledgments}
The work of JPL, HSS, YJZ is supported in part by CNRS via the LIA FCPPL. JPL is supported in part by the TMD@NLO IN2P3 project. The work of HSS is supported by the ILP Labex (ANR-11-IDEX-0004-02, ANR-10-LABX-63). YJZ is supported by the National Natural Science Foundation of China (Grants No. 11722539). NY was supported
by the JSPS Postdoctoral Fellowships for Research Abroad.
\end{acknowledgments}

\appendix

\section{Additional plots: further comparisons with data\label{app:moreplots}}

This appendix gathers additional plots of comparisons between our SPS calculations and experimental data collected by the ATLAS, CMS and LHCb experiments.

%
%

\subsection{Further comparisons with theory including partial CS LI corrections\label{app:LImoreplots}}

\begin{figure}[hbt!]
\centering
\foreach \page in {2}{
\includegraphics[page=\numexpr \page\relax, width=.32\textwidth,draft=false]{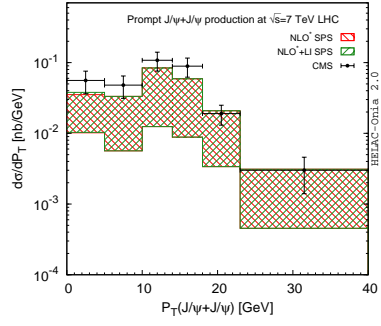}
}
\caption{$P_T(J/\psi+J/\psi)$ distribution for di-$J/\psi$ production via CS SPS within the CMS acceptance at $\sqrt{s}=7$ TeV ~\cite{Khachatryan:2014iia}.}\label{fig:cmslimore}
\end{figure}

\begin{figure}[hbt!]
\centering
\foreach \page in {2,3}{
\includegraphics[page=\numexpr \page\relax, width=.32\textwidth,draft=false]{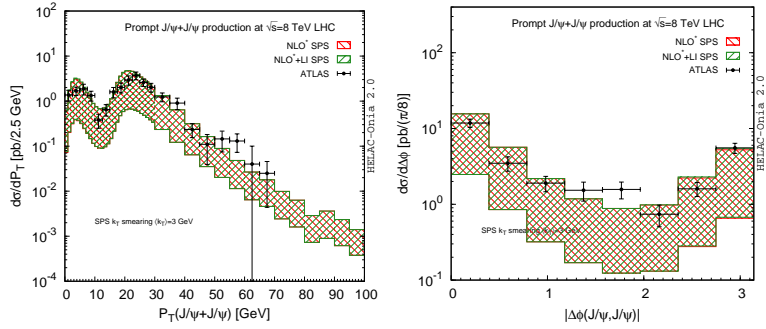}
}
\caption{$P_T(J/\psi+J/\psi)$ (left) and $\Delta \phi(J/\psi,J/\psi)$ (right) distributions for di-$J/\psi$ production via CS SPS within ATLAS $\sqrt{s}=8$ TeV acceptance~\cite{Aaboud:2016fzt}.}\label{fig:atlaslimore}
\end{figure}

\begin{figure}[hbt!]
\centering
\foreach \page in {1,3,4,5,6,7}{
\includegraphics[page=\numexpr \page\relax, width=.32\textwidth,draft=false]{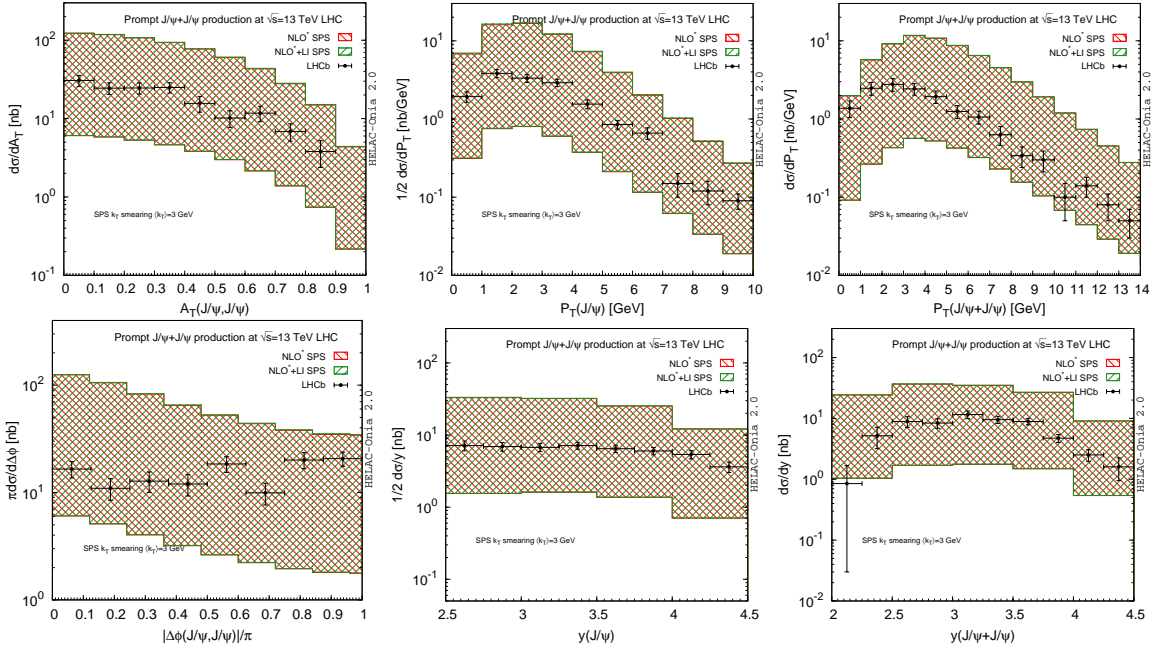}
}
\caption{$A_T(J/\psi,J/\psi)$ (top left), $P_T(J/\psi)$ (top middle), $P_T(J/\psi+J/\psi)$ (top right), $\Delta \phi(J/\psi,J/\psi)$ (bottom left), $y(J/\psi)$ (bottom middle) and $y(J/\psi+J/\psi)$ (bottom right) distributions for di-$J/\psi$ production via CS SPS within LHCb $\sqrt{s}=13$ TeV acceptance~\cite{Aaij:2016bqq}.}\label{fig:lhcblimore}
\end{figure}

We compare below our SPS CS NLO$^\star$+LI calculation to the experimental data for other observables than the rapidity difference and the invariant mass. The transverse-momentum distributions of the pair $P_T(J/\psi+J/\psi)$ are shown in Figure~\ref{fig:cmslimore} (CMS), in the left panel of Figure \ref{fig:atlaslimore} (ATLAS) and in the top-right panel of Figure \ref{fig:lhcblimore} (LHCb). The NLO$^\star$+LI SPS green bands almost overlay at the red bands (NLO$^\star$ SPS), which implies that these LI corrections are not important for these distributions. It is interesting to note that the initial $k_T$-smearing effect is important in the low $P_T(J/\psi+J/\psi)$ region, which illustrates that this distribution is indeed ideal to extract the transverse-momentum dependent information from the colliding partons inside the protons

\subsection{Further comparisons with theory including CO contributions\label{app:comoreplots}}

\begin{figure}[hbt!]
\centering
\foreach \page/\ipage in {2/1,9/2,8/3,4/4,1/5,3/6,6/7,7/8,5/9}{
\subfloat[Set \ipage]{\includegraphics[page=\numexpr \page\relax, width=.32\textwidth,draft=false]{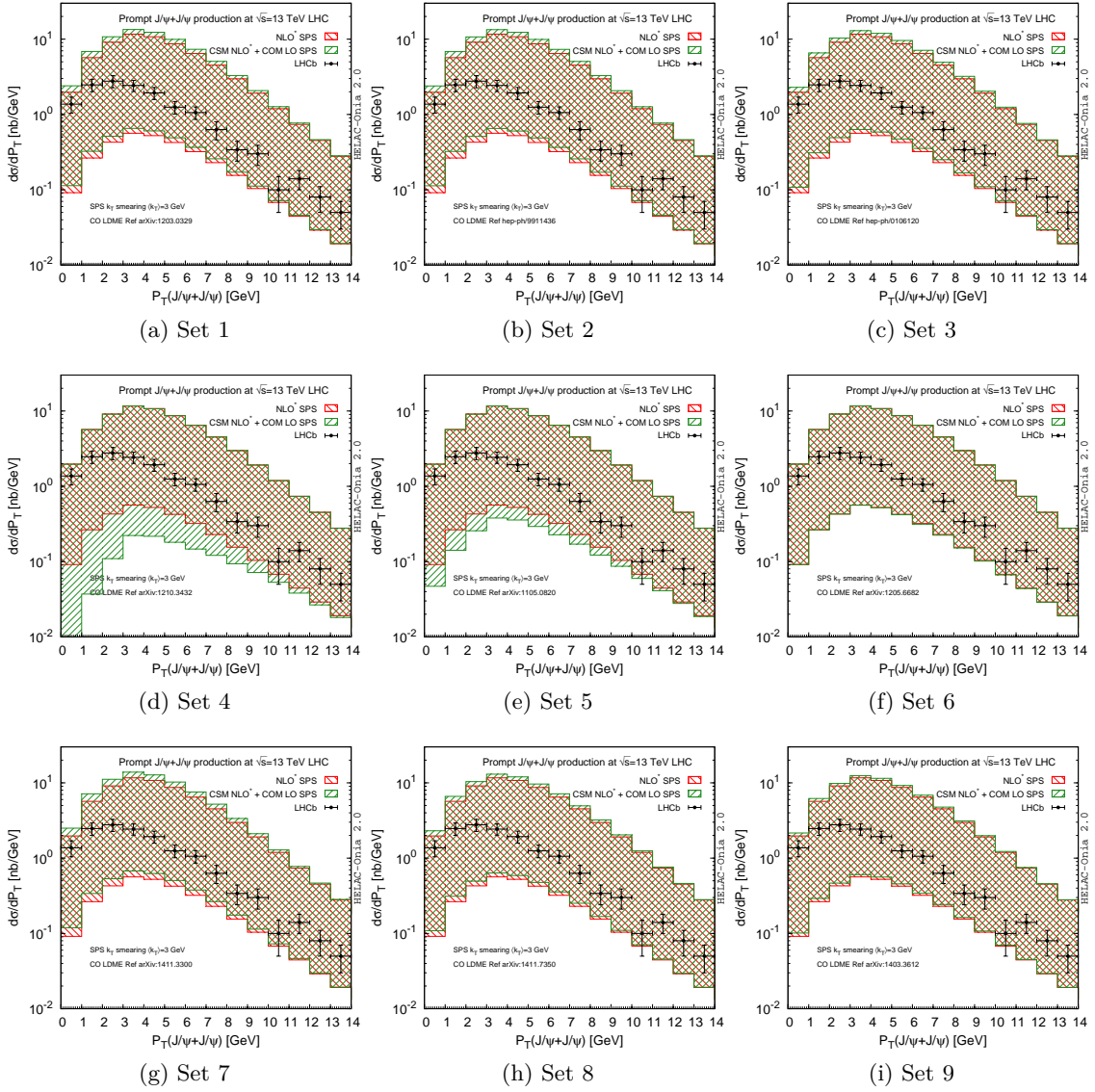}}
}
\caption{$P_T(J/\psi+J/\psi)$ distributions via SPS  NLO$^\star$ CS and LO CO  in the LHCb acceptance at $\sqrt{s}=13$ TeV~\cite{Aaij:2016bqq}.}\label{fig:colhcbdptpair}
\end{figure}

\begin{figure}[hbt!]
\centering
\foreach \page/\ipage in {2/1,9/2,8/3,4/4,1/5,3/6,6/7,7/8,5/9}{
\subfloat[Set \ipage]{\includegraphics[page=\numexpr \page\relax, width=.32\textwidth,draft=false]{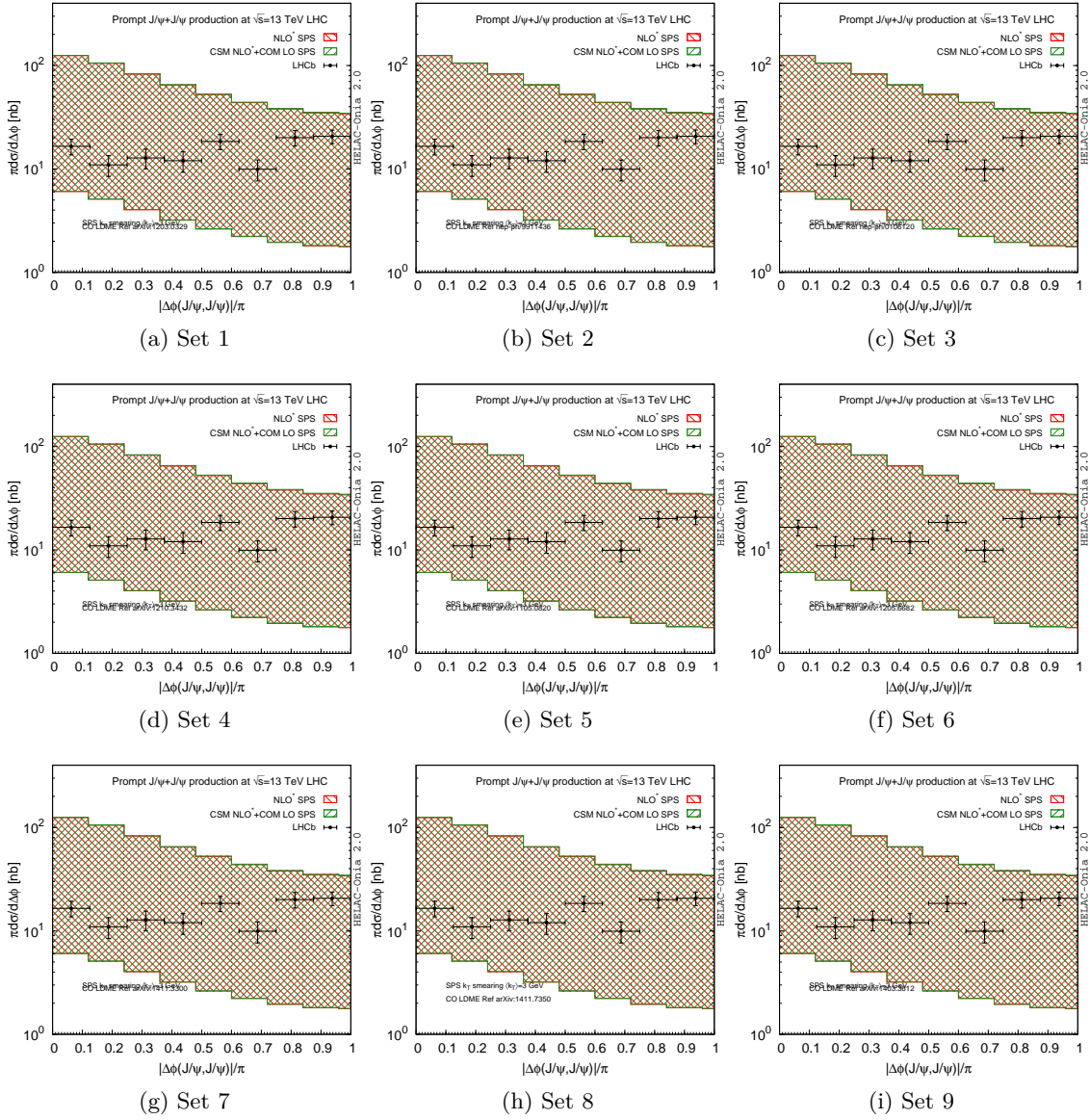}}
}
\caption{$\Delta \phi(J/\psi,J/\psi)$ distributions via SPS  NLO$^\star$ CS and LO CO  in the LHCb acceptance at $\sqrt{s}=13$ TeV~\cite{Aaij:2016bqq}.}\label{fig:colhcbdphi}
\end{figure}

\begin{figure}[hbt!]
\centering
\foreach \page/\ipage in {2/1,9/2,8/3,4/4,1/5,3/6,6/7,7/8,5/9}{
\subfloat[Set \ipage]{\includegraphics[page=\numexpr \page\relax, width=.32\textwidth,draft=false]{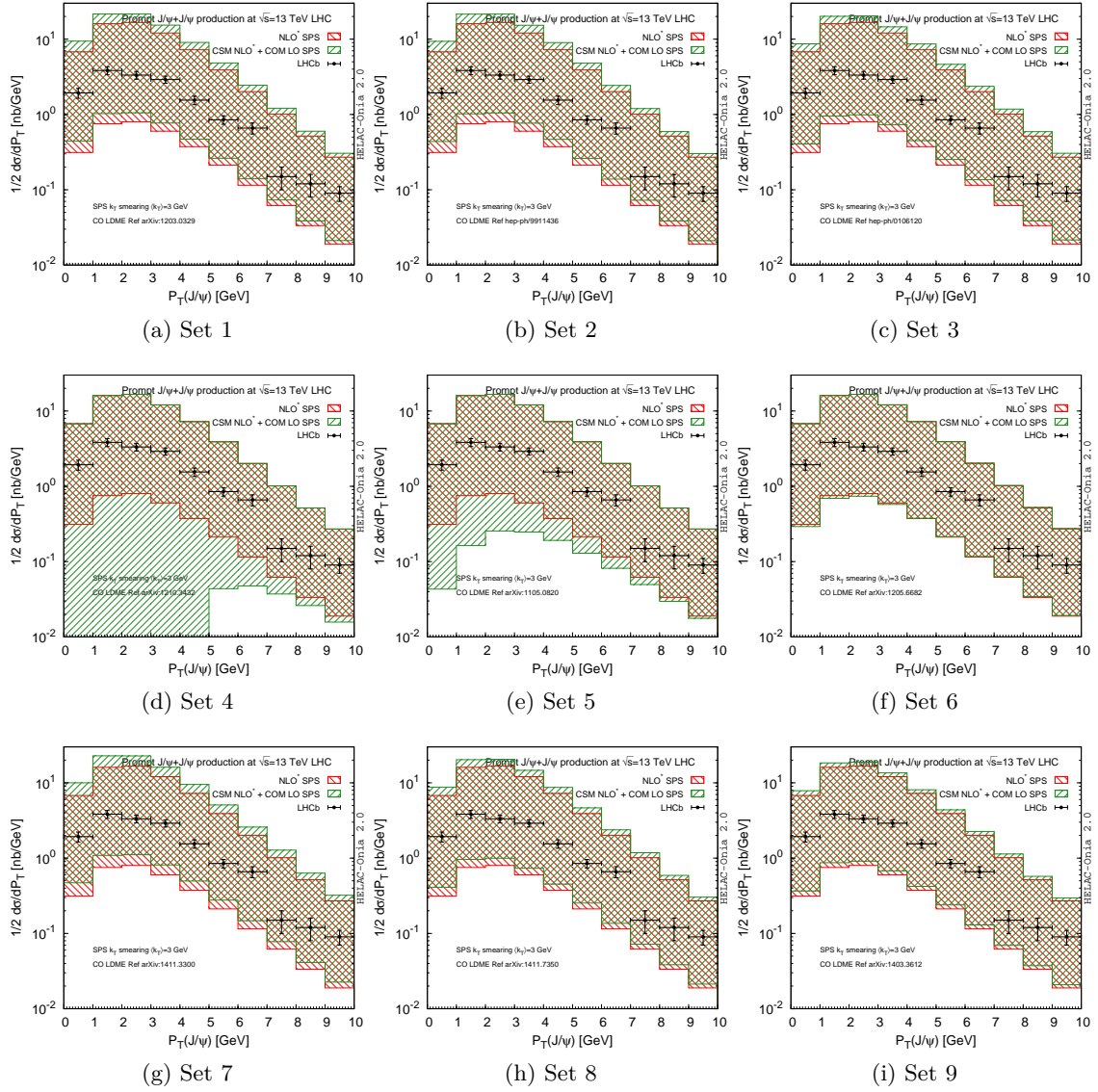}}
}
\caption{$P_T(J/\psi)$ distributions via SPS  NLO$^\star$ CS and LO CO  in the LHCb acceptance at $\sqrt{s}=13$ TeV~\cite{Aaij:2016bqq}.}\label{fig:colhcbdpt1}
\end{figure}

\begin{figure}[hbt!]
\centering
\foreach \page/\ipage in {2/1,9/2,8/3,4/4,1/5,3/6,6/7,7/8,5/9}{
\subfloat[Set \ipage]{\includegraphics[page=\numexpr \page\relax, width=.32\textwidth,draft=false]{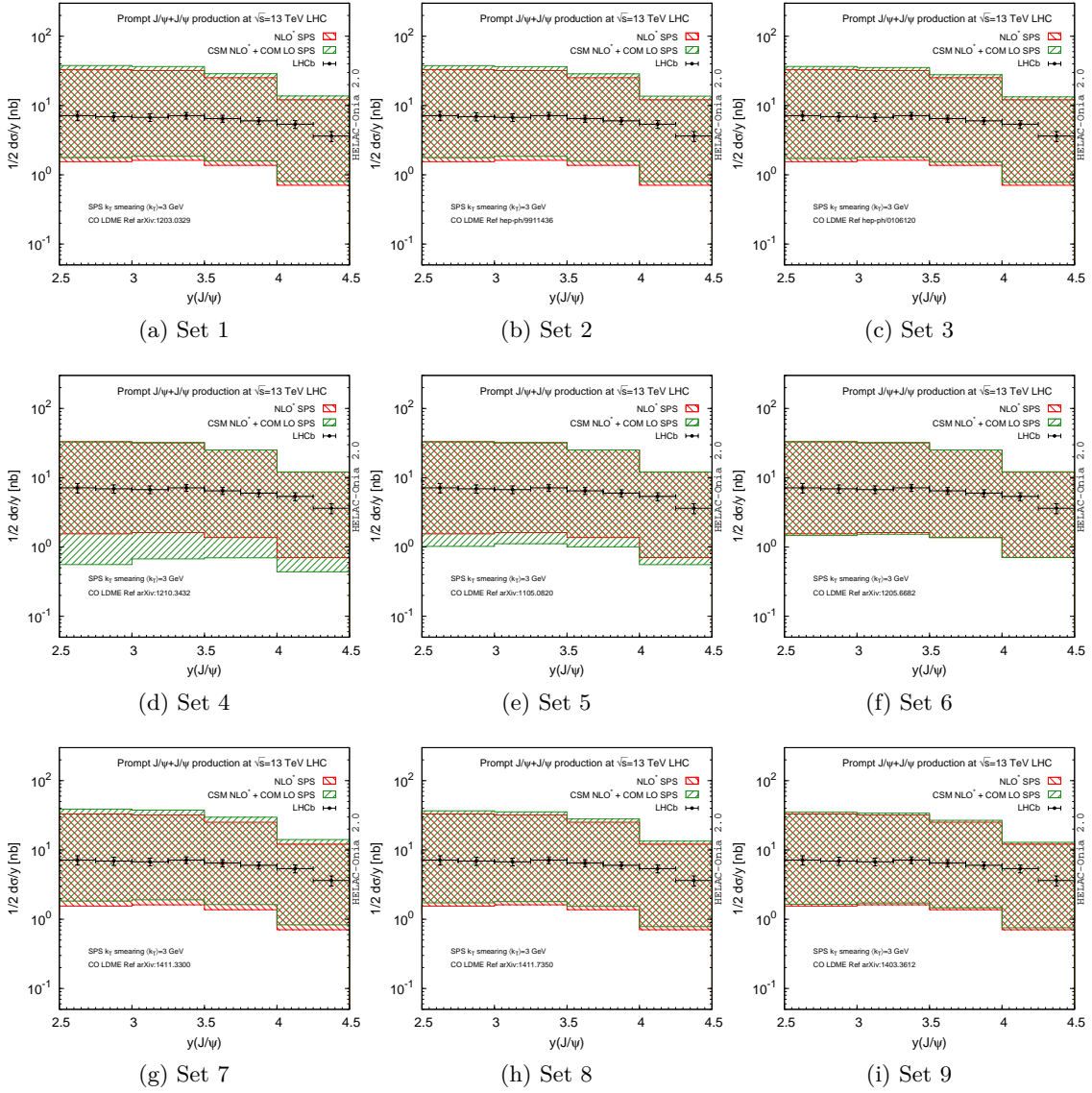}}
}
\caption{$y(J/\psi)$ distributions via SPS  NLO$^\star$ CS and LO CO  in the LHCb acceptance at $\sqrt{s}=13$ TeV~\cite{Aaij:2016bqq}.}\label{fig:colhcbdy1}
\end{figure}

\begin{figure}[hbt!]
\centering
\foreach \page/\ipage in {2/1,9/2,8/3,4/4,1/5,3/6,6/7,7/8,5/9}{
\subfloat[Set \ipage]{\includegraphics[page=\numexpr \page\relax, width=.32\textwidth,draft=false]{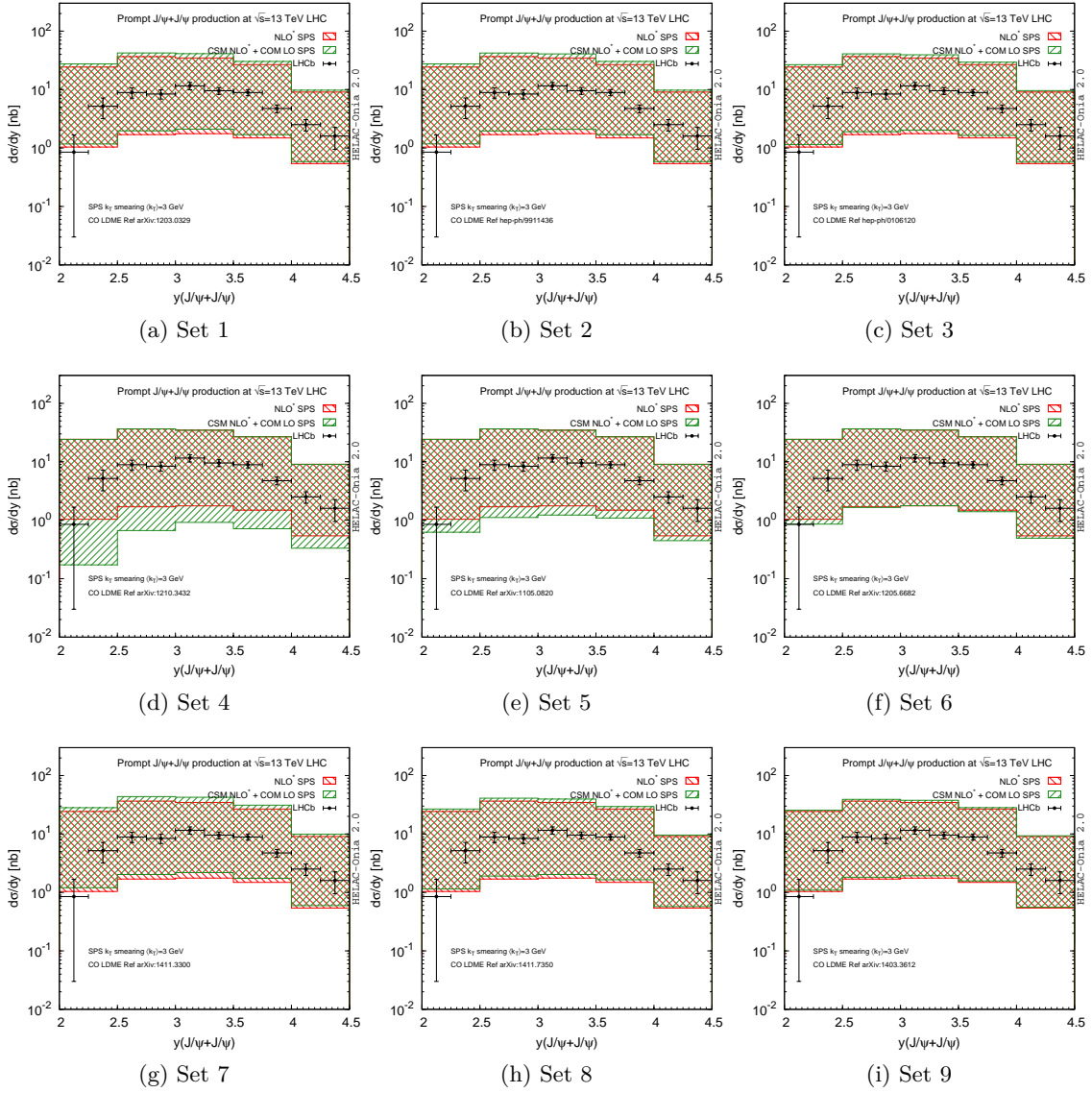}}
}
\caption{$y(J/\psi+J/\psi)$ distributions via SPS  NLO$^\star$ CS and LO CO  in the LHCb acceptance at $\sqrt{s}=13$ TeV~\cite{Aaij:2016bqq}.}\label{fig:colhcbdypair}
\end{figure}

\begin{figure}[hbt!]
\centering
\foreach \page/\ipage in {2/1,9/2,8/3,4/4,1/5,3/6,6/7,7/8,5/9}{
\subfloat[Set \ipage]{\includegraphics[page=\numexpr \page\relax, width=.32\textwidth,draft=false]{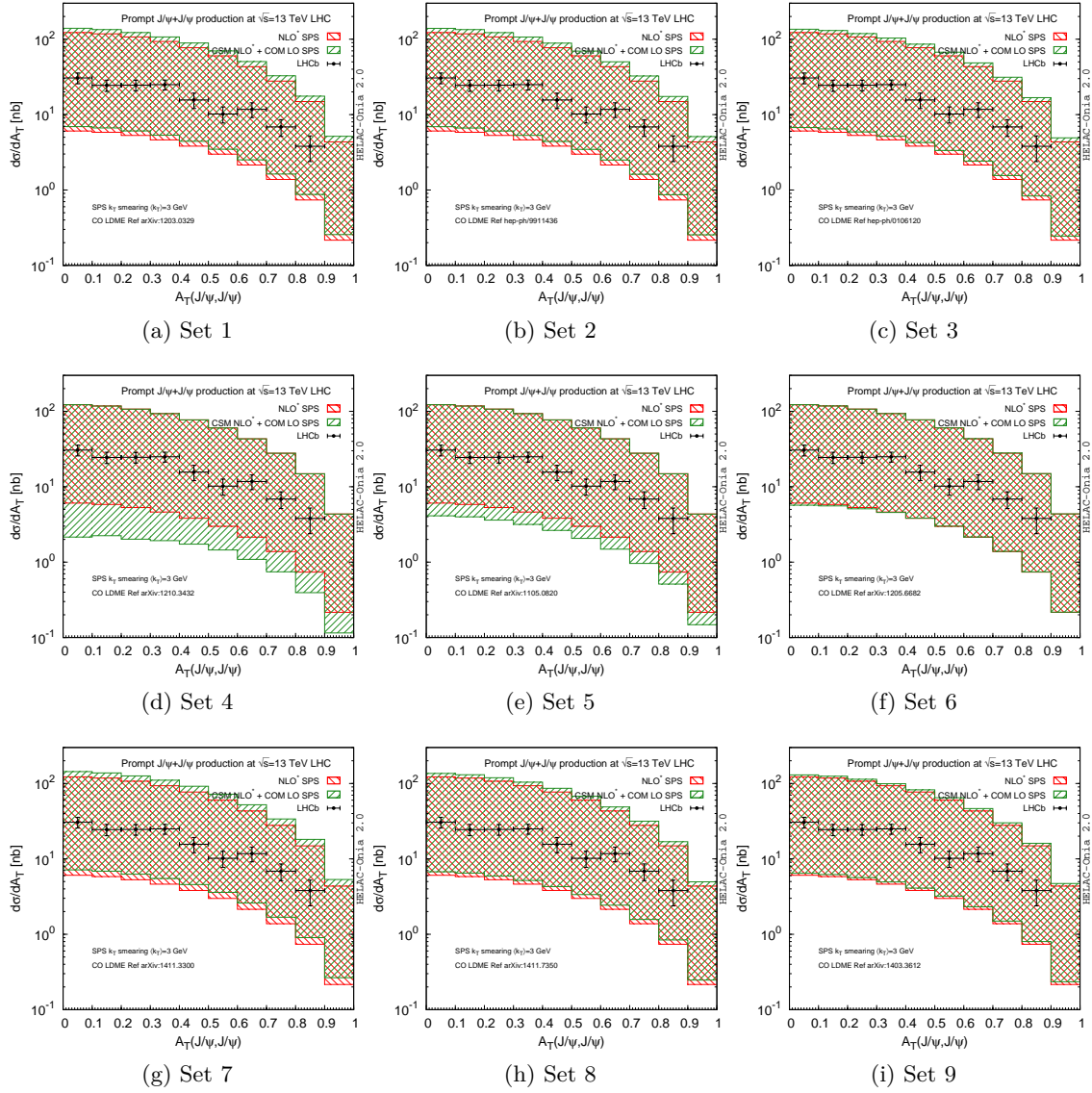}}
}
\caption{$A_T(J/\psi,J/\psi)$ distributions via SPS  NLO$^\star$ CS and LO CO  in the LHCb acceptance at $\sqrt{s}=13$ TeV~\cite{Aaij:2016bqq}.}\label{fig:colhcbdat}
\end{figure}

\begin{figure}[hbt!]
\centering
\foreach \page/\ipage in {2/1,9/2,8/3,4/4,1/5,3/6,6/7,7/8,5/9}{
\subfloat[Set \ipage]{\includegraphics[page=\numexpr \page\relax, width=.32\textwidth,draft=false]{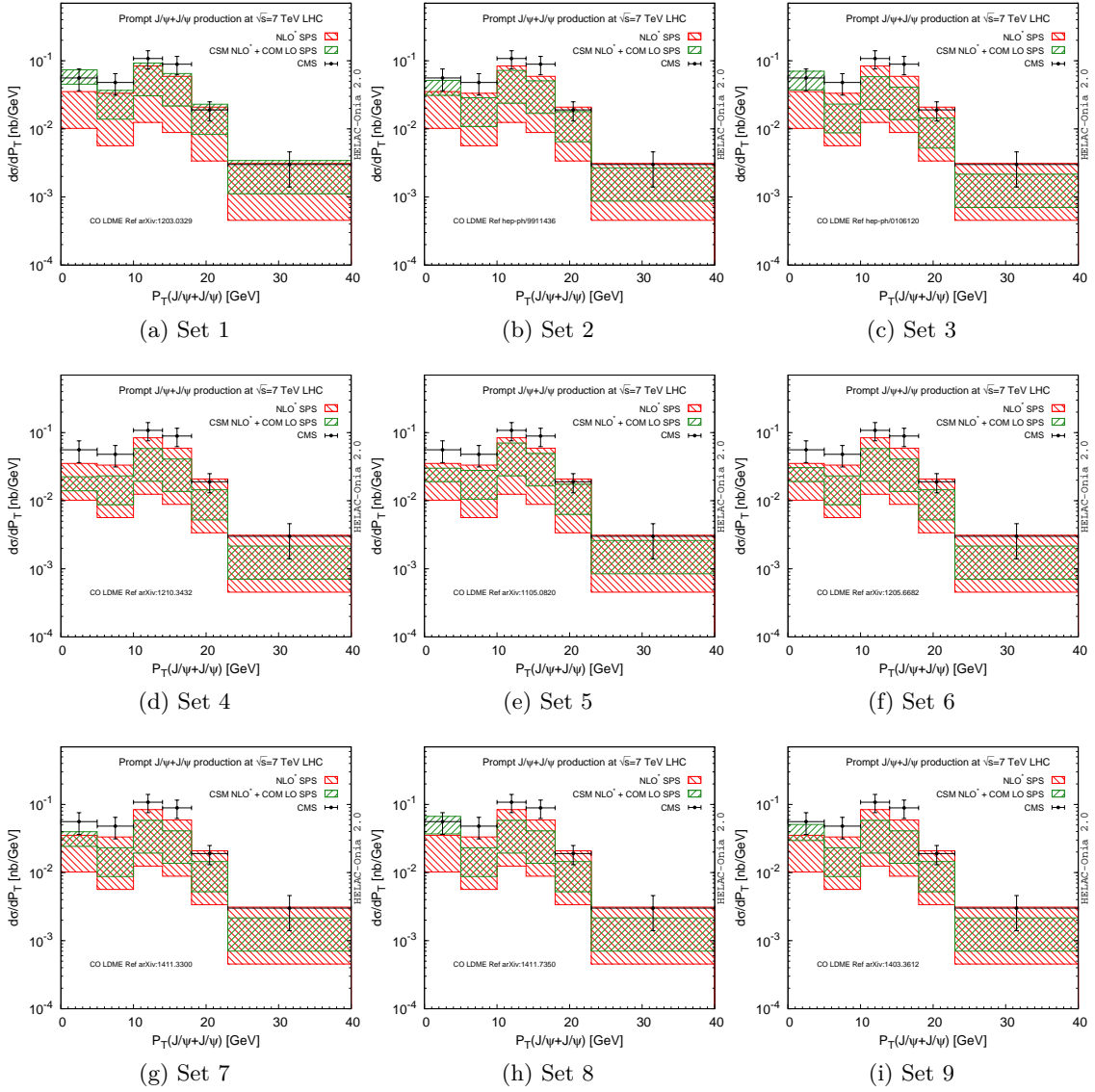}}
}
\caption{$P_T(J/\psi+J/\psi)$ distributions via SPS NLO$^\star$ CS and LO CO in the CMS acceptance at $\sqrt{s}=7$ TeV~\cite{Khachatryan:2014iia}.}\label{fig:cocmsdpt}
\end{figure}

\begin{figure}[hbt!]
\centering
\foreach \page/\ipage in {2/1,9/2,8/3,4/4,1/5,3/6,6/7,7/8,5/9}{
\subfloat[Set \ipage]{\includegraphics[page=\numexpr \page\relax, width=.32\textwidth,draft=false]{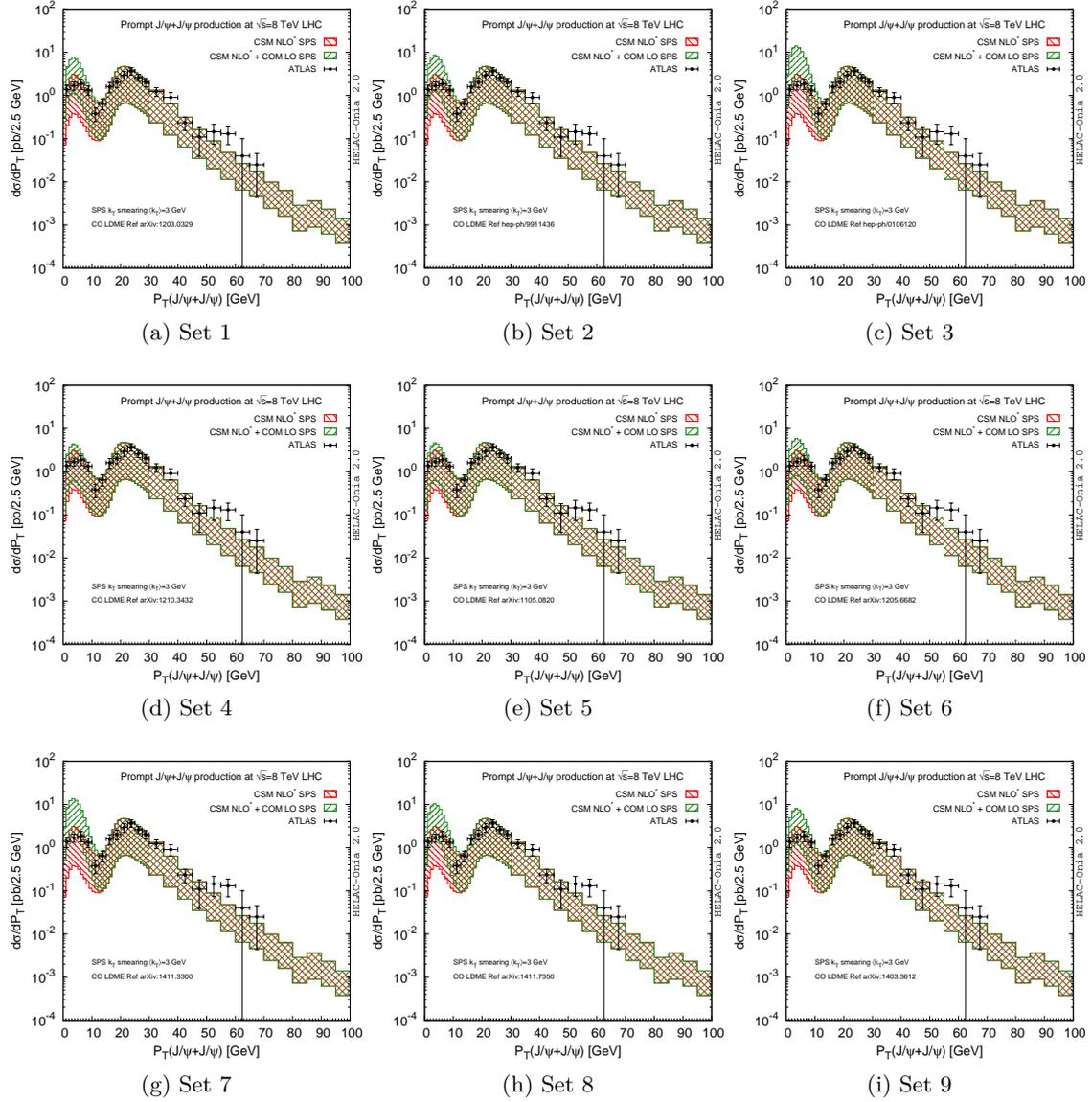}}
}
\caption{$P_T(J/\psi+J/\psi)$ distributions via SPS NLO$^\star$ CS and LO CO in the ATLAS acceptance at $\sqrt{s}=8$ TeV~\cite{Aaboud:2016fzt}.}\label{fig:coatlasdpt}
\end{figure}

\begin{figure}[hbt!]
\centering
\foreach \page/\ipage in {2/1,9/2,8/3,4/4,1/5,3/6,6/7,7/8,5/9}{
\subfloat[Set \ipage]{\includegraphics[page=\numexpr \page\relax, width=.32\textwidth,draft=false]{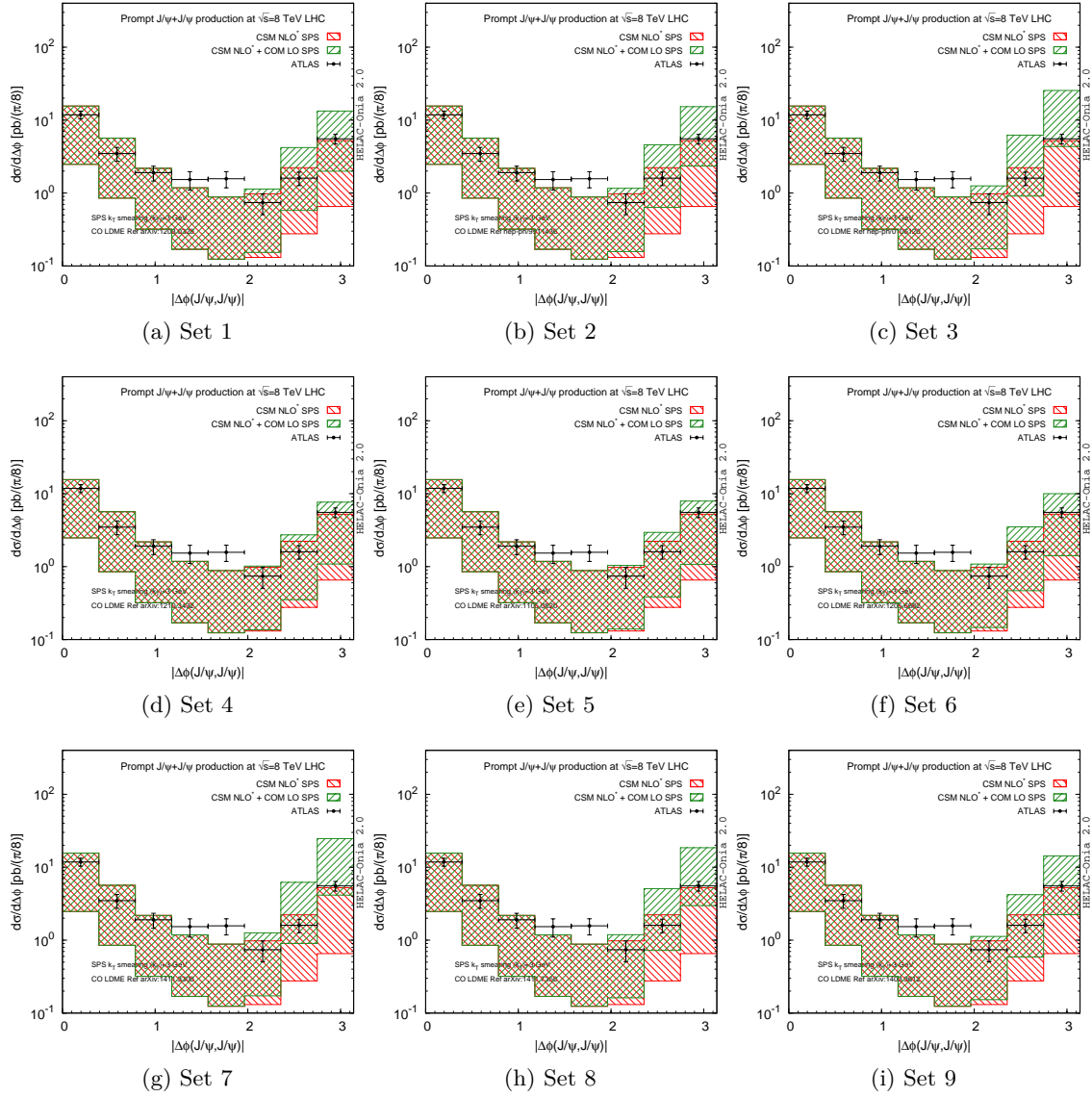}}
}
\caption{$\Delta \phi(J/\psi,J/\psi)$ distributions  via SPS NLO$^\star$ CS and LO CO in the ATLAS acceptance at $\sqrt{s}=8$ TeV acceptance~\cite{Aaboud:2016fzt}.}\label{fig:coatlasdphi}
\end{figure}

Further comparisons between CS NLO$^\star$+CO LO SPS results and LHCb data are shown in Figure~\ref{fig:colhcbdptpair} for $P_T(J/\psi+J/\psi)$, Figure \ref{fig:colhcbdphi} for $\Delta \phi(J/\psi,J/\psi)$, Figure \ref{fig:colhcbdpt1} for $P_T(J/\psi)$, Figure~\ref{fig:colhcbdy1} for $y(J/\psi)$, Figure \ref{fig:colhcbdypair} for $y(J/\psi+J/\psi)$ and Figure \ref{fig:colhcbdat} for $A_T(J/\psi,J/\psi)$ respectively. The inclusion of CO channels only slightly changes the corresponding predicted distributions of the SPS yield regardless of the set of LDMEs. Similar conclusions can be drawn for $P_T(J/\psi+J/\psi)$ and $\Delta \phi(J/\psi,J/\psi)$ distributions in the CMS and ATLAS acceptances, which is clearly seen in Figures~\ref{fig:cocmsdpt}, \ref{fig:coatlasdpt} and \ref{fig:coatlasdphi}.

\bibliographystyle{utphys}
\bibliography{paper}
\end{document}